\documentclass[10pt, journal,compsoc]{IEEEtran}
\usepackage[T1]{fontenc}
\usepackage{xcolor}
\usepackage{float}
\usepackage{cite}  
\usepackage{enumerate}
\usepackage{url}

\usepackage{breakurl}
\usepackage[breaklinks]{hyperref}
\usepackage{bookmark}
\usepackage{array}
\newcommand{\HDY}[1] 
{\textcolor{black}{#1}}

\ifCLASSINFOpdf
   \usepackage[pdftex]{graphicx}
\else
\fi
\begin{document}
%
\title{A Survey on Visualization Approaches in Political Science for Social and Political Factors: Progress to Date and Future Opportunities}
%
%
%
%
\author{Dongyun Han, 
Abdullah-Al-Raihan Nayeem, 
Jason Windett, Yaoyao Dai, Benjamin Radford,
\\ and Isaac Cho, \textit{Member, IEEE}

\IEEEcompsocitemizethanks{
\IEEEcompsocthanksitem
 Dongyun Han and Isaac Cho are with Utah State University. \protect E-mail: \{dongyun.han, isaac.cho\}@usu.edu
\IEEEcompsocthanksitem
 Abdullah-Al-Raihan Nayeem, Jason Windett, Yaoyao Dai, and Benjamin Radford are with the University of North Carolina at Charlotte. \protect  E-mail: \{anayeem, jwindett, yaoyao.dai, Benjamin.radford\}@uncc.edu
}

\thanks{Manuscript received April 19, 2005; revised August 26, 2015.}}

%
%

\markboth{Journal of \LaTeX\ Class Files,~Vol.~14, No.~8, August~2015}%
{Shell \MakeLowercase{\textit{et al.}}: Bare Demo of IEEEtran.cls for Computer Society Journals}
%



\IEEEtitleabstractindextext{%
\begin{abstract}
Politics is the set of activities related to strategic decision-making in groups. Political scientists study the strategic interactions between states, institutions, politicians, and citizens; they seek to understand the causes and consequences of those decisions and interactions. While some decisions might alleviate social problems, others might lead to disasters such as war and conflict. 
\HDY{Data visualization approaches have the potential to assist political scientists in their studies by providing visual contexts.}
However, political researchers’ perspectives on data visualization are unclear. This paper examines political scientists' perspectives on visualization and how they apply data visualization in their research. We discovered a growing trend in the use of graphs in political science journals. However, we also found a knowledge gap between the political science and visualization domains, such as effective visualization techniques for tasks and the use of color studied by visualization researchers. 
To reduce this gap, we survey visualization techniques applicable to the political scientists’ research and report the visual analytics systems implemented for and evaluated by 
political scientists. 
At the end of this paper, we present an outline of future opportunities, including research topics and methodologies, for multidisciplinary research in political science and data analytics. Through this paper, we expect visualization researchers to get a better grasp of the political science domain, as well as broaden the possibility of future visualization approaches from a multidisciplinary perspective.
\end{abstract}

}

\maketitle

\IEEEdisplaynontitleabstractindextext

%
\IEEEpeerreviewmaketitle


%
%
%
%

\IEEEraisesectionheading{\section{Introduction}\label{sec:introduction}}

\begin{figure*}[t]
    \centering
    \includegraphics[width=.95\textwidth]{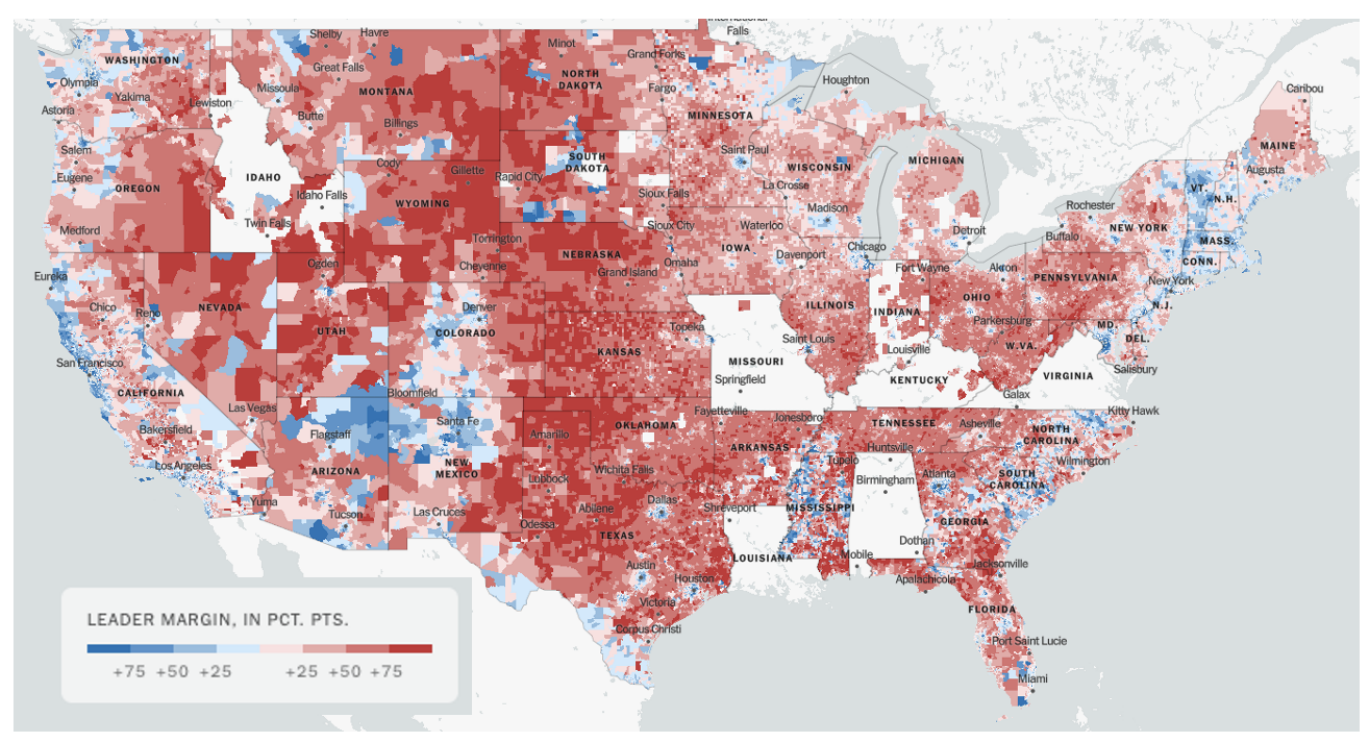}
    \caption{An Extremely Detailed Map of the 2020 Presidential Election from New York Times. It shows votes data from 2,523 of 3,143 counties in 47 U.S. states.
    }
    \label{extream_map}
\end{figure*}

\IEEEPARstart{D}{ata} visualization is more than just the presentation of data in graphical form; it can support political scientists in their reasoning and decision-making processes by providing visualized contexts, patterns, and trends~\cite{dork2013critical, traunmuller2020visualizing, segel2010narrative}. This makes sense given that politics is closely related to data visualization's history - visualization methods have evolved with the introduction of new social, political, and economic data collections~\cite{friendly2001milestones}.
As an introductory example, consider an election for a president or prime minister of a country. Data visualization approaches provide many types of statistical graphs, such as choropleth maps, cartograms, and line and bar charts that depict various social and political indicators including political ideology, policy salience, and vote percentages across states or districts (e.g., ``An Extremely Detailed Map of the 2020 Election''~\cite{election_nyt} (Fig. \ref{extream_map})). These approaches could help researchers and the general audience contextualize election-related data to gain better situational awareness and forecast electoral outcomes.

Data visualization researchers 
have long studied how people interpret interactive graphics and naturally assimilate visual information. As data visualization visually transforms different types of data in various domains, visualization techniques are diversified. It has been demonstrated that visualization can support the complex tasks of domain experts. 
\HDY{Political science} is no different. 
For example, Cao et al.~\cite{cao2012whisper} and Wang et al.~\cite{wang2012si} introduced novel visualization tools for political scientists. They report that their proposed systems successfully supported political scientists' tasks (i.e., understanding social media or public opinion and analyzing network connections) through a variety of evaluation methods including usage scenarios, user studies, and expert feedback. However, in a curious way, only a few approaches are studied from the perspective of collaboration and interdisciplinary research between the fields of political science and visualization compared to the importance of politics and public interest~\cite{zinovyev2010data, kastellec2007using}.  

The objective of this report is to provide a structured overview for visualization researchers to better understand the tasks of political scientists and their analysis requirements. 
We first review visualization techniques for political scientists' analysis goals. Following that, we report on the requirements for the design of tasks and systems using cognitive task taxonomies~\cite{amar2005low}. 
Our review reveals that relatively few visualization methods have been proposed for political scientists.

Next, data visualization perspectives from political scientists are discussed. Although political science journals have recently encouraged the use of graphs,  
we only found limited visualization approaches. We attribute this to the different viewpoints on data visualization from political scientists. It could also aid visualization researchers in understanding the views of political scientists on data visualization, with the hope of fostering interdisciplinary understanding and collaboration between the fields of political science and data visualization. 

This paper is organized as follows. In Section~\ref{sec_review_necessity}, we first explore the closeness between the field of data visualization and political science and argue the necessity of this work. Next, Section~\ref{sec_scope_methodology} presents our survey methodologies that describe the scope and method of collection, as well as taxonomies. Subsequently, we review visualization approaches in Section~\ref{sec_review_vis} and \ref{sec_review_vis_system} and report the perspectives on data visualization in the field of Political Science in Section \ref{sec_perspective_political_science}. 
Finally, we discuss the knowledge gap between researchers in two fields and report the conclusion to encourage multidisciplinary collaborations in Section~\ref{sec_knowledgegap} and ~\ref{sec_conclusion}.
\subsection{Necessity of Review}\label{sec_review_necessity}

The history of the data visualization field can be traced back to social science. Visualization methods have evolved with the introduction of new social, political, and economic data collections~\cite{friendly2001milestones}. For example, basic graphical formats such as line, bar, and pie charts were first introduced by William Playfair, a Scottish political economist, to visualize political, economic, and social data~\cite{playfair1801commercial}. 
Edward Tufte, one of the most famous figures in modern data visualization, whose work focuses on conveying quantitative information to broad audiences through visualization, is also a political scientist ~\cite{tufte:2001}. 
However, political scientists' perspectives on data visualization have changed significantly over time. Although political science journals are reporting on the growing trend of displaying data in graphical format, political scientists still tend to prefer tables and infographics to interactive data visualization ~\cite{traunmuller2020visualizing, kastellec2007using}. 
This has potentially created a knowledge gap between visualization and political scholars on data visualization.

We found three papers that introduce and encourage data visualization approaches in the political science fields \cite{zinovyev2010data, kastellec2007using, traunmuller2020visualizing}. 
\HDY{Traunmüller~\cite{traunmuller2020visualizing}, and Kastellec and Leoni~\cite{kastellec2007using} review top political science journals, which are discussed in Section~\ref{sec_perspective_political_science}. }
They review visualization techniques such as matrix, scatter plot, and map, and present their usage examples and advantages in political science. 
However, these papers focus only on using data visualization as a reporting tool, \HDY{not as an analysis tool.} 


Compared to previous review papers, our report mainly focuses on visualization approaches designed for political scientists in visualization venues. 
This report provides not only an overview of visualization techniques but also visualization applications along with their domain-specific tasks and design considerations. 
In addition, this paper examines political scientists' perspectives on data visualization. In general, domain experts are generally satisfied with their conventional methods for their data analysis processes~\cite{batch2017interactive}. Similarly, political scientists may have little motivation to change their practices, even though visualization approaches have advanced and been studied by incorporating knowledge from other academic fields such as cognitive science and statistics. 
Our work could serve as a promotion of the visualization approach for political scientists and facilitate interdisciplinary understanding and collaboration with political scientists and visualization researchers.




\section{Survey Scope and Methodology}\label{sec_scope_methodology}


\subsection{Political Scientists}

We first clarify who \textit{political scientists} are for this survey. Political scientists are a subset of social scientists who broadly study political processes and public policy at the local, state, national, and global levels. 
Political scientists are social scientists primarily concerned with understanding how political actors wield power.
\HDY{\textit{Political actors} can be broadly defined} to be entities involved in policy making or other political processes that include but are not limited to, politicians, government officials, \HDY{(non-) profit organizations}, political consultants, public administrators, media, and citizens. 
Political actors, for example, negotiate, strategize, and compromise in order to determine which social issues are prioritized and what policy options are available and preferable for addressing those issues. 
In addition, political scientists often also serve as consultants or advisors to political actors in their particular areas of expertise.
This may involve, for example, advising campaigns, helping to craft effective policies, or providing expertise on the political processes of foreign countries. \HDY{Please note that the other political actors besides political scientists are outside the scope of our target visualization users in this work. }

\subsection{\HDY{Literature Collection \& Scope}}

We create a database for the data visualization literature comprising a collection of peer-reviewed articles published in nine journals and conferences on visualization, graphics, and system domains by 2022. The journals and conferences are listed in Table~\ref{tab:source_vis}. They are reviewed to investigate and discuss visualization approaches designed to assist political scientists. 

\begin{table}
    \scriptsize
    \begin{center}
    \begin{tabular}{|p{1cm}|p{7cm}    |}
        \hline
        TVCG & IEEE Transactions on Visualization and Computer Graphics \\
        \hline
        CG\&A & IEEE Computer Graphics and Applications \\
        \hline
        TiiS & ACM Transactions on Interactive Intelligent Systems  \\
        \hline
        TIST & ACM Transactions on Intelligent Systems and Technology \\
        \hline
        VIS & IEEE Visualization Conference \\
        \hline
        CGF & Computer Graphics Forum\\
        \hline
        CHI & ACM Conference on Human Factors in Computing Systems \\
        \hline
        IUI & ACM Conference on Intelligent User Interfaces \\
        \hline
        \end{tabular}  
    \label{tab:source_vis}
    \end{center}
    \caption{ Journals \& Conferences for Visualization, Graphics, and Intelligent System  }
\end{table}

To populate our database for the visualization field, we sought to collect papers from major data visualization journals and proceedings designed to support political scientists. Making a decision about which papers are suitable for our survey by inspecting the titles, abstracts, and paper keywords (i.e., Association for Computing Machinery (ACM) Computing Classification System) is subtle and not trivial. To collect our literature data set, we conduct the following steps. First, we collect literature in the target journals and proceedings beginning in the year 2000. Some of the venues began after 2000, so the first years of the paper collections from each venue are different. Next, we collect bib files for each venue and target year. 
In the case of IEEE VIS, we use the visualization publication dataset~\cite{isenberg2016vispubdata} that contains the papers published between 1990 and 2020. We access a download page for each paper using the Digital Object Identifier (DOI) information or Uniform Resource Locator (URL) described in the files. We write a Python script to automatically collect each paper using Selenium. We only manually download the papers published in IEEE VIS 2021 and 2022 because their bib sources are not available. 


After we collect the visualization literature, we conduct three literature selection processes. 
First, we prepare a set of unigrams to check whether each paper contains the target keywords. The general keyword terms include `policy', `politics', `census', `demographic', `government', and `governor.' 
We also use the following keywords which describe \HDY{subfields of political science including \textit{International Relationships}, \textit{Comparative Politics}, and \textit{American Politics}~\footnote{\url{https://en.wikipedia.org/wiki/Category:Subfields_of_political_science}}}: `election', `congression', `presidency', `conflict', `war', `terrorism', `anarchy', `democracy', `socialism', `governance', `democracy', `autocracy', `authoritarianism', `populism', `revolution', `regime', `censorship', `repression.' These keywords are provided by the three political scientists in the author list. 

\begin{figure}[t]
    \centering
    \includegraphics[width=.5\textwidth]{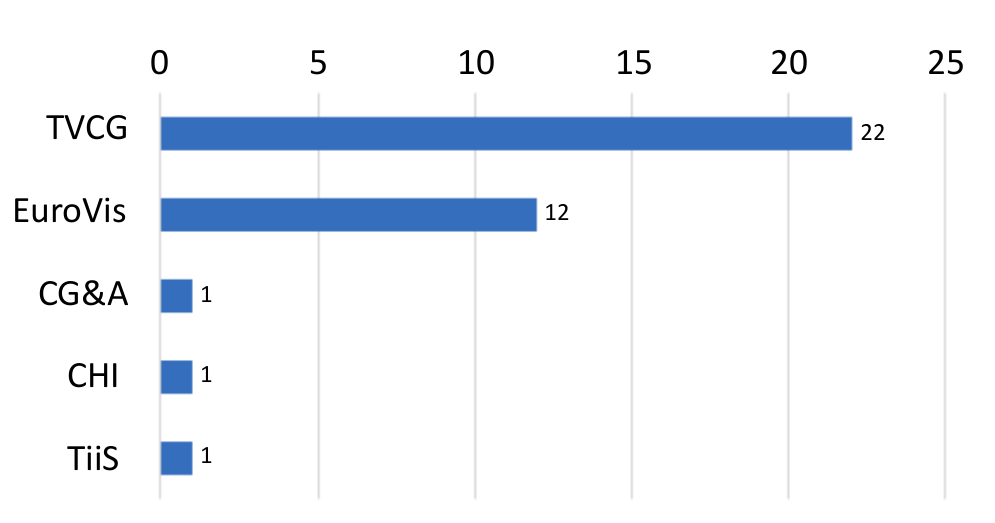}\\
    \caption{
    A total of 37 publications proposed visualization techniques and systems to support political scientists' tasks. 
    }
    \label{fig_vis_publication_counts}
\end{figure}

Next, we perform automatic keyword filtering by writing a Python script and then conduct manual secondary filtering of the literature containing the keywords. This is because words such as `demographic' and `policy' could be used with various meanings. 

Finally, we select the literature that introduces visual analytic tools that are collaborated with and designed for political scientists. 
\HDY{We find lots of research works in fields such as urban planning and traffic management, which could be indirectly relevant to political issues. Nevertheless, those works fall under the following cases would not be considered for inclusion in our survey.
}
Many research papers published in the \HDY{journals and conferences of CHI, TiiS, TIST, and IUI} discuss how to encourage citizens to participate in public politics, e-Governance, and community policies and introduce web systems to \HDY{encourage community participation~\cite{hsu2017community}} and evaluate user behaviors~\cite{wan2014improving}.
\HDY{However, most of these works are not relevant to data visualizations.}
\HDY{Next,} we find multiple visualization systems designed for political actors such as policymakers and government agencies. For example, EpiPolicy~\cite{tariq2021planning} and Epinome~\cite{livnat2012epinome} are designed for institutional actors in the Department of Public Health to develop and carry out complex infection control plans for the prevention of various mass-casualty diseases such as COVID-19, Ebola, and SARS. \HDY{We decided not to cover these papers since the purpose of this work is to investigate data visualization approaches for political scientists, and the other political actors are beyond the scope of our target users. } 
We also discover that social media analytics piqued the interest of visualization researchers, some of whom used public and political opinions to demonstrate the effectiveness of their methods. 
Among those articles, we only collect those in which visualization researchers collaborated with political scientists to design visualization approaches because visualizations for social media analytics are already discussed \HDY{in earlier works~\cite{chen2017social, wu2016survey}.}
As a result, the remaining final number of articles is 37 and is shown in Figure~\ref{fig_vis_publication_counts} by the venues.



\subsection{Taxonomy of Survey}

\begin{figure}[t]
    \centering
    \includegraphics[width=.5\textwidth]{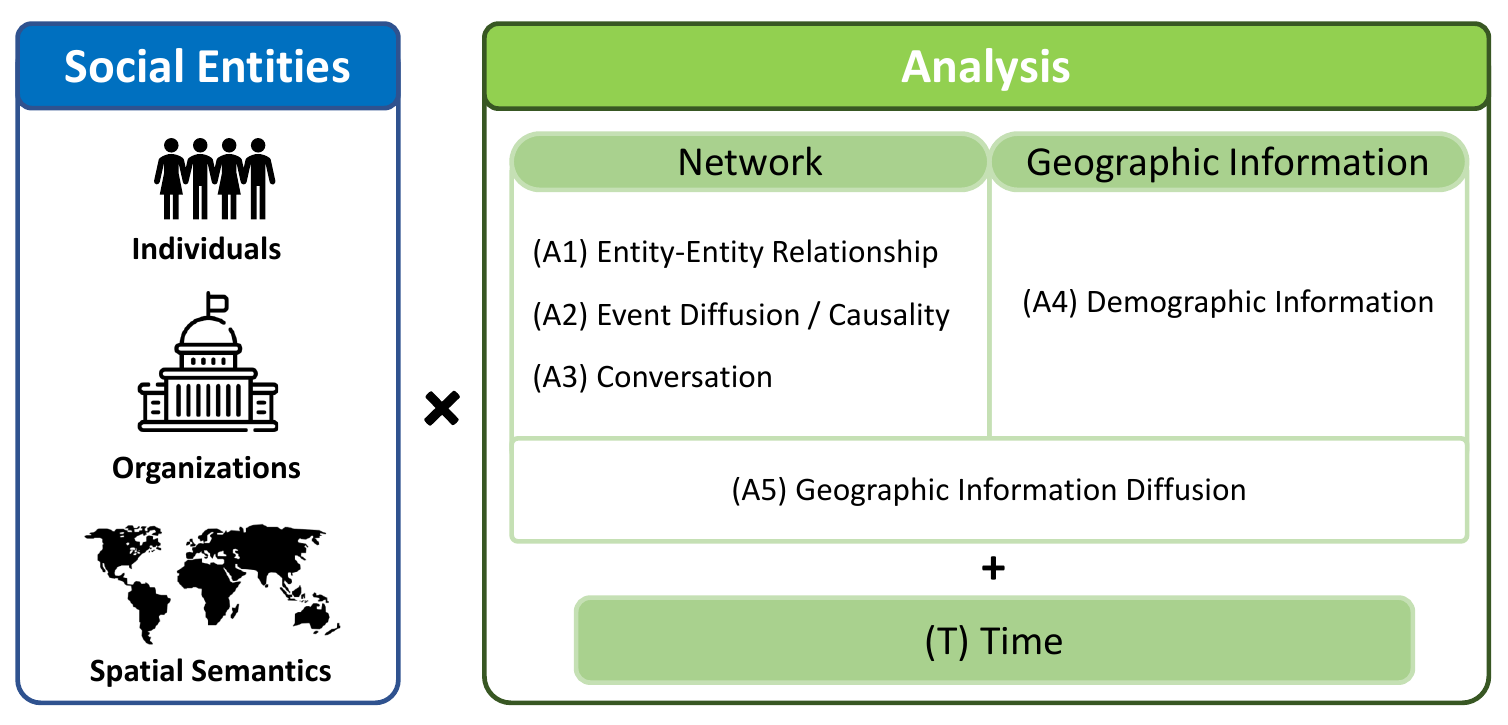}
    \caption{Social entities include individuals, organizations, and spatial semantics (e.g., city, state, and country levels). Political scientists are interested in political systems and how social entities interact with each other within different political systems. We found the 5 analysis goals (i.e., A1-A5) which are mainly discussed in the visualization venues.
    }
    \label{fig_vis_entities_goals}
\end{figure}

To review visualization approaches, the following taxonomies are used. 

\textbf{\HDY{Domain Analytical Goal Taxonomy.}} Political scientists are mainly interested in analyzing relationships between social entities. 
\HDY{From the 37 research papers, we find five main analytical purposes of visual analytics approaches to support domain analysis. 
The five domain analytical goals are \textit{A1. Entity-Entity Relationship, A2. Event Diffusion/Causality, A3. Conversation, A4. Demographic Information,} and \textit{A5. Geographic Information Diffusion}} as shown in Fig.~\ref{fig_vis_entities_goals}. In Section~\ref{sec_review_vis}, we review visualization techniques along with these five domain analytical goals.

\textbf{\HDY{Visual Analytical Task Taxonomy.}} 
\HDY{Understanding which analytical tasks could be supported by visualization techniques is important for visualization researchers in designing visualization systems.
In Section~\ref{sec_review_vis_system}, we review visualization applications by domain interests and analyze their design requirements by visual analytical tasks.} 
To examine this, we use seven task types out of ten introduced by Amar et al.~\cite{amar2005low} as summarized in Table~\ref{tab:task_type_taxonomy}. \HDY{The other three analytical tasks (\textit{Find Extremum, Sort,} and \textit{Determine Range}) are not used in this work. 
This is because, while we find the visualization systems in Section~\ref{sec_review_vis_system} support these functions, they are not specified in their design requirements.
}


\begin{table}
    \scriptsize
    \begin{center}
    \begin{tabular}{|p{1.8cm}|p{6.2cm}    |}
        \hline
        \textbf{Task Type} & \textbf{Description} \\ 
        \hline
        Retrieve Value & Given a data case, find attributes of those cases \\
        \hline
        Filter & Given some concrete conditions, find data cases satisfying those conditions \\
        \hline
        {Compute \newline Derived Value} & Given a data case, compute an aggregate numeric representation of those data cases \\
        \hline
        Characterize Distribution & Given a data case and a quantitative attribute of interest, characterize the distribution of that attribute's values \\
        \hline
        Find \newline Anomalies &  Identify any anomalies within a given data case with respect to a given relationship or expectation \\
        \hline
        Cluster & Given a data case, find clusters of similar attribute values \\
        \hline
        Correlate & Given a data case and two attributes, determine useful relationships between those attributes\\
        \hline
        \end{tabular}
    \end{center}
    \label{tab:task_type_taxonomy}
    \caption{Task Type Taxonomy: tasks and their descriptions are defined by Amar et al.~\cite{amar2005low} }
    \vspace{-.5cm}
\end{table}

\begin{table*}[t!]
\begin{center}
    \includegraphics[width=\textwidth]{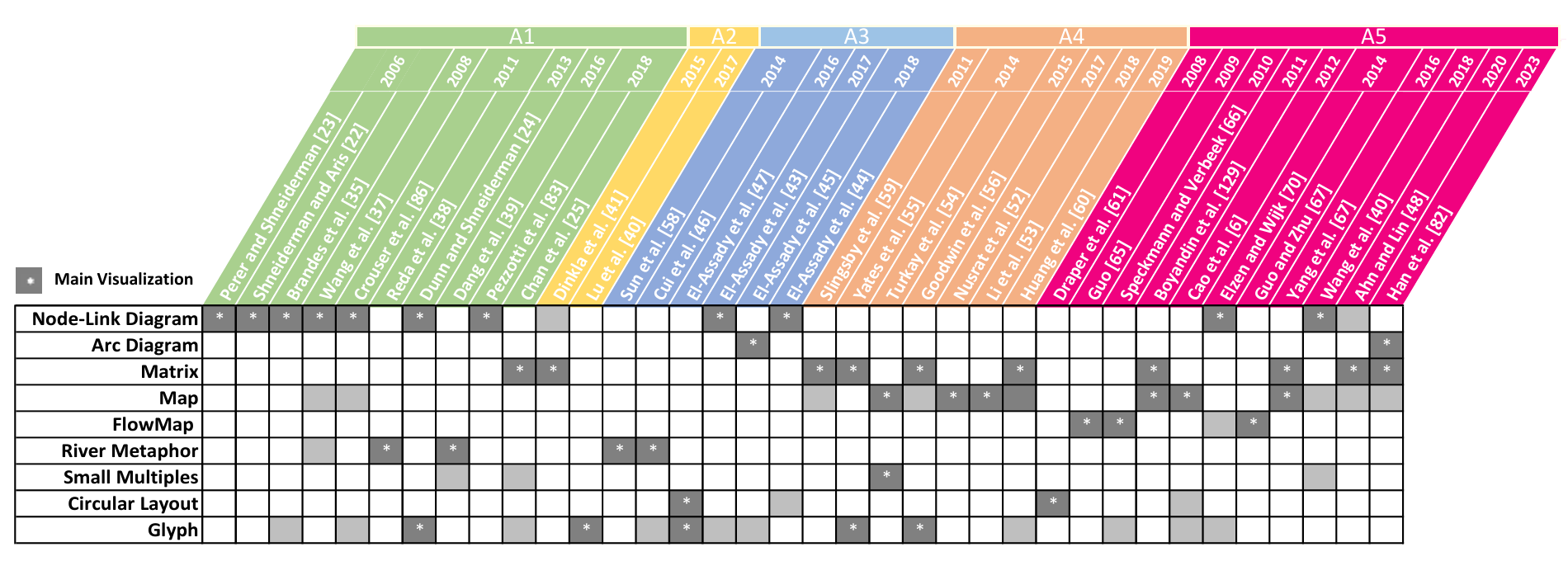}\\
    \caption{36 out of the 37 selected papers on visualization and visual analytics for political scientists. More than one visualization could be used in a visualization work. They are colored by analysis goals. The dark gray color with a star indicates the main visualization technique.  The light gray colors denote visualizations used for contextual information rather than the 5 analysis goals. The excluded paper is Cashman et al.~\cite{cashman2020cava}, a visual interface for building data structure. }
    \label{fig_visualization_category}
    \end{center}
    \vspace{-.5cm}
\end{table*}

\begin{figure*}[t]
\begin{center}
    \includegraphics[width=\textwidth]{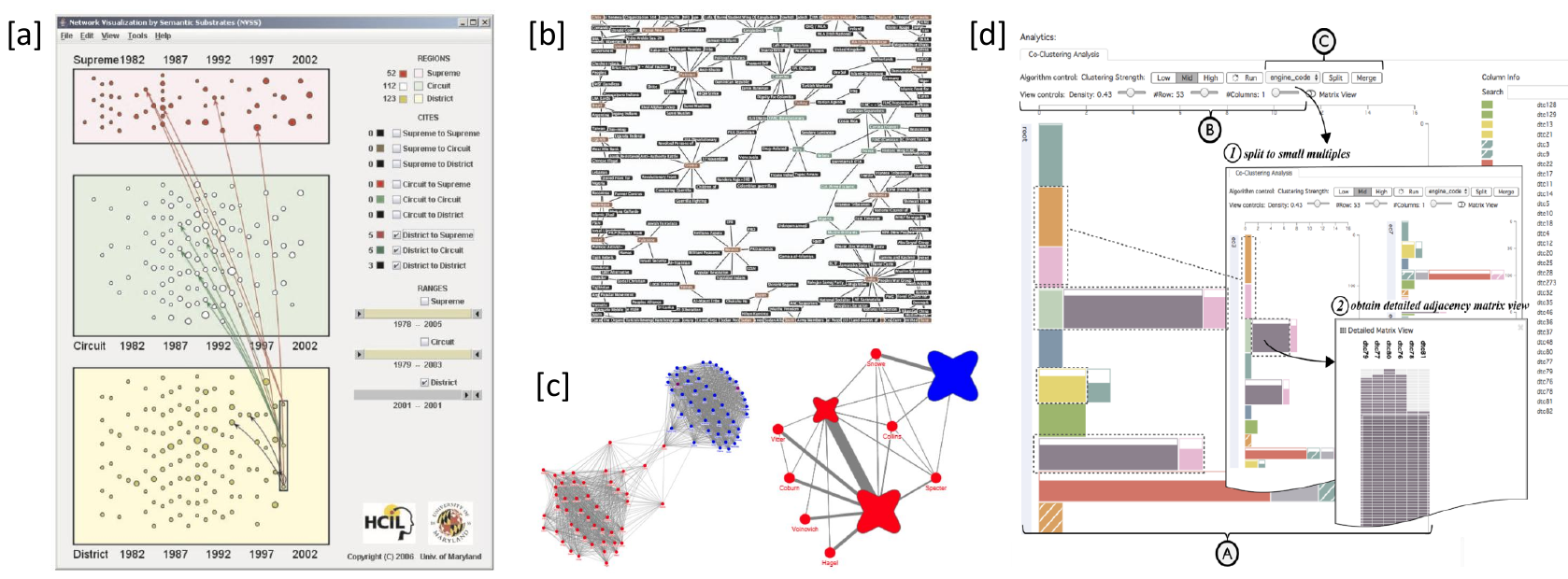}\\
    \caption{Network visualizations for entity relations analysis (\textbf{a}: Schneiderman and Aris~\cite{shneiderman2006network}, \textbf{b}: SocialAction~\cite{perer2006balancing}, \textbf{c}: Motif Simplification~\cite{dunne2013motif}, \textbf{d}: VIBR~\cite{chan2018v}). 
    }
    \label{fig_entityRelation_network}
    \end{center}
    \vspace{-.5cm}
\end{figure*}

\section{Visualization Techniques}\label{sec_review_vis}
This section examines the visualization techniques according to the \HDY{five domain analytical goals.}
For each goal, we review the related visualization techniques. 
Its summary is presented in Table~\ref{fig_visualization_category}. The background colors of the authors indicate the main category (A1-A5). Research papers may have more than one visual element like glyphs and small multiples. The darker gray color with a white star indicates the main visualization category of each work, while the light gray color means additional visual elements. We sub-categorize them by the main visualization.

\subsection{Datasets}
In political science, datasets are usually network or hierarchical and could be structured or unstructured. Structured data is data sources that are already cleaned and preprocessed, such as government-released statistics. Example datasets are congressional vote records, domestic policy diffusion records~\cite{SPID_dataset}, global trade network~\cite{FAOSTAT} and indicators~\cite{WGI, theWorldBank}, and global terror events~\cite{raleigh2010introducing}. Finding such reliable datasets is important. This could be difficult to search without collaborating with domain experts (e.g., political scientists). 
However,  thanks to social and political researchers and governors, many relevant datasets are publicly available online. Existing data repositories for political and social data sets are Harvard Dataverse~\cite{HarvardDataverse}, the University of Michigan Library~\cite{michigan_finding_data, michigan_finding_statistics}, and the U.S. Government's open data ~\cite{data_gov}.

Political scientists also collect and analyze rich (semi-) unstructured datasets comprising, such as news articles, speeches, and social media network data. They are frequently voluminous and noisy. Natural Language Processing (NLP) techniques, such as Named-Entity Recognition (NER), are often utilized to extract social entities as well as spatiotemporal and numerical information. The social entities include individuals, organizations, and spatial semantics (i.e., city, state, and nation). They are the target subjects in whom political scientists are interested (Fig.~\ref{fig_vis_entities_goals}).



\subsection{Entity Relationship Analysis (A1)}
Entity relationship analysis is objective to understanding social interactions between political actors. 
As political actors are connected globally and relevant data size grows, it is getting more difficult to understand how social entities are related to each other within a system~\cite{brandes2006summarizing}. 

\textbf{Network Visualization.} 
\HDY{Network representations are suitable for investigating structural relationships. Such entity relation graphs could be large, so reducing visual clutter is very important. } Schneiderman and Aris~\cite{shneiderman2006network} introduce a semantic approach for node placement to address a visual clutter problem. As shown in Fig.~\ref{fig_entityRelation_network}a, it has divided cluster spaces according to the properties of nodes vertically. Individual nodes are secondarily placed in each cluster space considering not to be overlapped with each other. This helps users cope with the complexities of large numbers of nodes, while links can only be seen after a user has chosen nodes. They claim that their method effectively helps political science researchers comprehend precedent-setting patterns like whether Supreme Court cases rely more heavily on lower courts over time. Perer and Shneiderman~\cite{perer2006balancing} introduce SocialAction, pointing out that many analysts use network visualizations for opportunistic exploration rather than orderly data navigation. 
It prioritizes nodes having higher centrality metrics, and the prioritized nodes are located close to the center as shown in Fig.~\ref{fig_entityRelation_network}b. Utilizing network centrality is a common practice these days to provide an overview, identify important nodes or outliers, and filter nodes~\cite{shen2006visual}. Its usability is demonstrated with terrorist groups committing over 70,000 terrorist attacks across the world during a 27-year period.

Reducing visual clutter is also available by grouping nodes having similar patterns. Dunne and Shneiderman~\cite{dunne2013motif} introduce motif simplification (Fig.~\ref{fig_entityRelation_network}c). It allows visualization practitioners to save screen space, while it is advantageous for users to quickly understand underlying network structures. It is designed by political scientists, and its power is evaluated with the 2007 U.S. Senate voting pattern network. 

Another challenge for Network representations is to present relationships with geographic information. 
Wang et al.~\cite{wang2008investigative} plot terror events according to terrorist organizations as points and link them in chronological order. \HDY{It is a simple practice and helps political scientists establish and understand the geographical patterns and distribution of terrorist activities. }

\textbf{Matrix Visualization.} A matrix visualization enables users to explore relationships at a glance by placing entities or attributes in rows and columns and by encoding colors to cells along with them. For example, PDViz (Fig.~\ref{fig_pdviz}) aligns two blue and red rectangle cells vertically within one square cell across the U.S. states and political topics. The blue and red intensities represent the relative number of policy creations and existing policy adoption by each topic respectively. \HDY{It allows users to identify which U.S states are political leaders and have a greater interest in a political topic than other states.}

VIBR~\cite{chan2018v} presents a novel adjacency list style visualization for bipartite relations (Fig.~\ref{fig_entityRelation_network}d). 
It is designed for large scale relations by clustering similar entities in two disjoint sets. Higher priority groups are placed on the left. Users could identify and understand bipartite relationships quickly by scanning the leftmost vertical items. 
Its usability is shown with congress voting analysis. It could provide an overview of the bipartite relations (i.e., Republicans and Democrats) and their roles in legislation processes. Compared to node-link diagrams, flow maps, and adjacency matrices, its visual design results are well aligned and compact.

\begin{figure}[t]
\begin{center}
    \includegraphics[width=1.0\columnwidth]{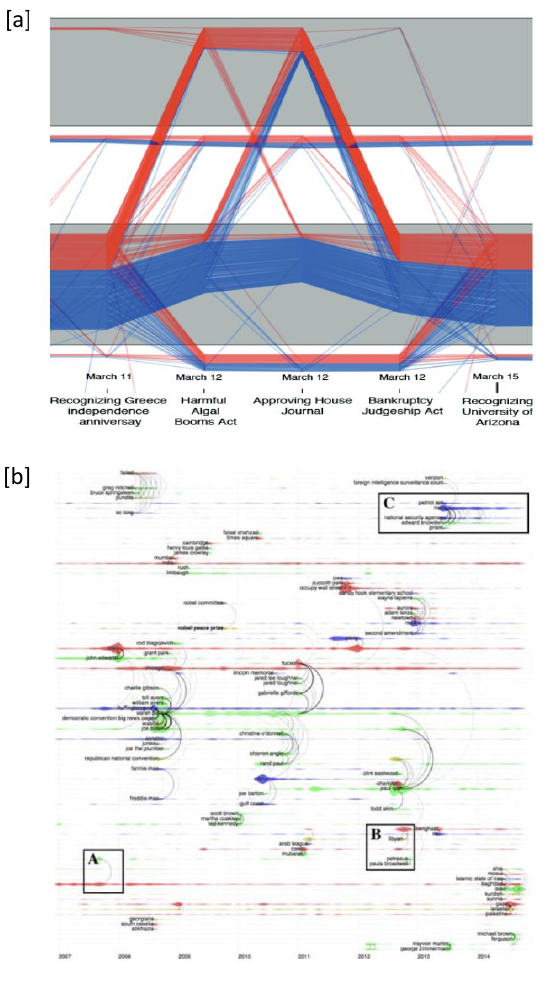}
    \caption{River metaphor visualizations for dynamic entity relation analysis over time (\textbf{a}: Reda et al.~\cite{reda2011visualizing}, \textbf{b}: TimeArcs~\cite{dang2016timearcs}). 
    }
    \label{fig_entityRelation_riverlike}
    \end{center}
    \vspace{-.5cm}
\end{figure}

\textbf{River Metaphor Visualization.}
A river metaphor visualization is used to address the challenge to visualize gradual relation changes over time. Reda et al.~\cite{reda2011visualizing} introduce a novel visualization approach that lists lines representing entities in a community. A distance between two lines at a given instant indicates how close two entities are to one another. As shown in Fig.~\ref{fig_entityRelation_riverlike}a, entity relations are changed over time. In congressional voting record data, for instance, the lines could represent individual politicians in the political parties (i.e., Republican, Democrat, and Independent). It allows users to analyze temporal patterns of dynamic entity relationships. Dang et al.~\cite{dang2016timearcs} also introduce a dynamic entity relation visualization, called TimeArcs (Fig.~\ref{fig_entityRelation_riverlike}b).
It aligns related entities close together leveraging force-directed layouts, making it easier for users to detect clusters within implicit network relations. The arcs show the relationships between two entities. It allows users to quickly identify relational changes over time. For instance, using 90,811 pieces of content from seven different Internet political media sources written over a ten-year period, TimeArcs demonstrated how political actor relationships dynamically change over time.

\subsection{Event Correlation / Causality Analysis (A2)}
Political disasters and tragedies do not happen alone, nor do they happen suddenly. Understanding event correlation is important for political scientists to make sense of a large number of events and identify the few events that stand out among the rest. This could be accomplished by close looking for and analyzing relationships between events \HDY{and social entities.}


 \textbf{Network Visualization.} 
 Wang et al.~\cite{wang2018visual} use a directed acyclic graph network representation (Fig.~\ref{fig_global_trade}) to visualize the potential flow of an anomaly within a global trade network. Nodes stand for countries in the graph. The leftmost node indicates the source of an anomaly with abrupt changes in global trade, and its connected countries are thought to be those that could be affected by the anomaly. The overall system of Wang et al.'s approach is designed for political scientists to analyze correlations between global trade anomalies and domestic stability.

\begin{figure}[t]
\begin{center}
    \includegraphics[width=0.5\textwidth]{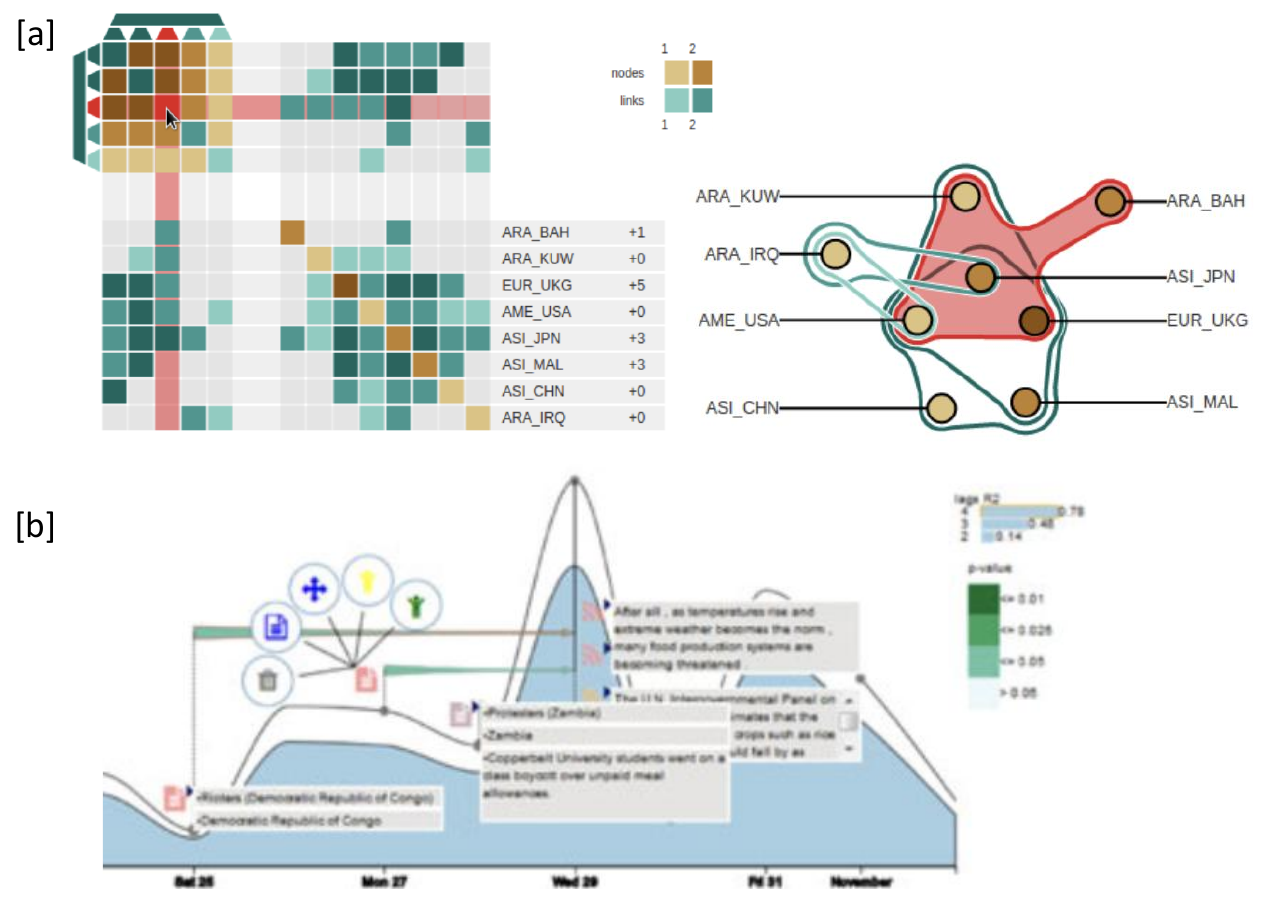}\\
    \caption{Visualizations for event correlation and causality. (a) Link-centric matrix and node-centric network visualization~\cite{dinkla2015dual}. (b) Line chart with glyphs showing event causalities~\cite{lu2017visual}.
    }
    \label{fig_correlation_matrix}
    \end{center}
    \vspace{-.5cm}
\end{figure}

\textbf{Matrix Visualization.} Grouping nodes and links is an efficient method to provide high-level structures of entity relationships. Dinkla et al.~\cite{dinkla2015dual} introduce Dual Adjacency Matrix (DAM), a link-centric matrix visualization. DAM supports users to do link-centric tasks such as finding common nodes shared by a given link group and finding link groups that share nodes. As shown in Fig.~\ref{fig_correlation_matrix}a, Dinkla et al. demonstrate that DAM has potential advantages when it is used with node-centric visualizations together. As a case study, they show that users can identify countries with similar patterns in global trade networks. However, it has been reported by domain experts that the use of DAM requires a steep learning curve.

\begin{figure*}[t]
\begin{center}
    \includegraphics[width=0.95\textwidth]{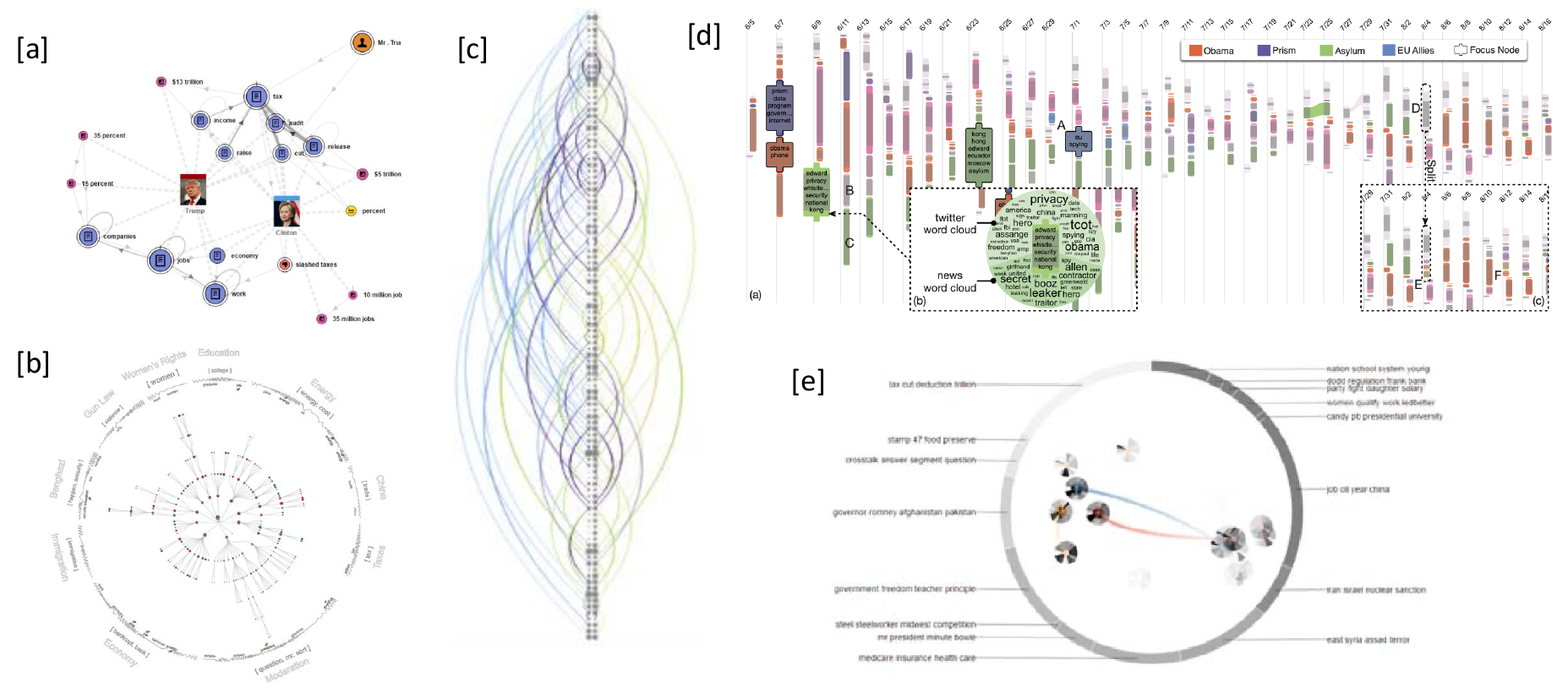}\\
    \caption{Visualizations for conversation analysis (\textbf{a}: NEREx~\cite{el2017nerex}, \textbf{b}: El-Assady et al.~\cite{el2018visual}, \textbf{c}: ThreadReconstructor~\cite{el2018threadreconstructor}, \textbf{d}: RoseRiver~\cite{cui2014hierarchical}, \textbf{e}: ConToVi~\cite{el2016contovi}).
    }
    \label{fig_conversation}
    \end{center}
\end{figure*}

PolicyFlow~\cite{ahn2020policyflow} introduces a novel metrics-style visualization, called the policy-inspection view (Fig.~\ref{fig_poliyflow}f). It shows how the policy was propagated from state to state in accordance with a selected policy. In the matrix view, the vertical and horizontal axes indicate the U.S. state and the adoption year respectively. On the y-axis, the U.S. states are listed in the order that they adopted the chosen policy, from bottom to top. Each cell encodes an individual adoption case or a measured policy diffusion pattern to indicate in which state created and adopted the policy. Users can determine whether each adoption case fits the pattern by comparing them all in one visualization.

\textbf{Glyph.} Lu et al.~\cite{lu2017visual} collaborate with experts from political science to design a media exploration tool with respect to armed conflict. They introduce a visualization that combines a line chart and glyphs to present causal relationships between events. It is shown in Fig.~\ref{fig_correlation_matrix}b. 
The line chart shows the volume of a media stream over time for a chosen media topic, and the gray dots represent related events that occurred on a specific date. The interface Lu et al. proposed has a causality test function. 
If two events have a causal relationship, an arrow glyph is drawn between them, and their significance is mapped from light green to dark green. The causality test model performance is represented by a colored area below the black media streamline.
According to Lu et al., their line plus glyph method enables their domain expert partner to quickly acquire the base knowledge of causality events and lead them to formalize hypotheses.

\subsection{Conversation Analysis (A3)}
Political scientists analyze multilateral discussions that have political, economic, and social relevance, such as political debates and oral court arguments \HDY{among political actors}. They are interested in exploring the social dynamics underlying dialogue, the role of speakers in the discussion process, the thematic evolution of discussion, and various argumentation strategies~\cite{habermas1985theory,persson2013effects}.

\textbf{Network Visualization.} El-Assay et al.~\cite{el2017nerex} propose NEREx. They extract a total of eight entities (e.g., people, objects, locations, etc) plus two positive and negative sentiments from conversation using NER to create an entity-relationship model network. They conduct a qualitative study with three domain experts in political science using 2016 US Presidential Debate data. Fig.~\ref{fig_conversation}a shows its result. The experts evaluate that NEREx helps them explore and comparatively analyze the debate. 

While these analytical approaches can be also performed using a series of political debates or presidential speeches, \HDY{political scientists must first restructure dialogues to detect and analyze their implicit structures.}
To visually support these processes, El-Assady et al.~\cite{el2018threadreconstructor} present ThreadReconstructor. It consists of two vertical arc diagrams as shown in Fig.~\ref{fig_conversation}c. Domain experts in political science evaluate that it enables them to quickly rebuild underlying conversation models from multiple dialogues. It also allows them to compare different dialogue models to create the best appropriate model using the human-in-the-loop approach. 
Similarly, El-Assady et al.~\cite{el2018visual} introduce a circular tree visualization for users to adopt their domain knowledge in the topic modeling processes to build better hierarchical topic results. Fig.~\ref{fig_conversation}b shows the topic modeling result from the second presidential debate between Romney and Obama in 2012.



\begin{figure*}[t]
\begin{center}
    \includegraphics[width=0.95\textwidth]{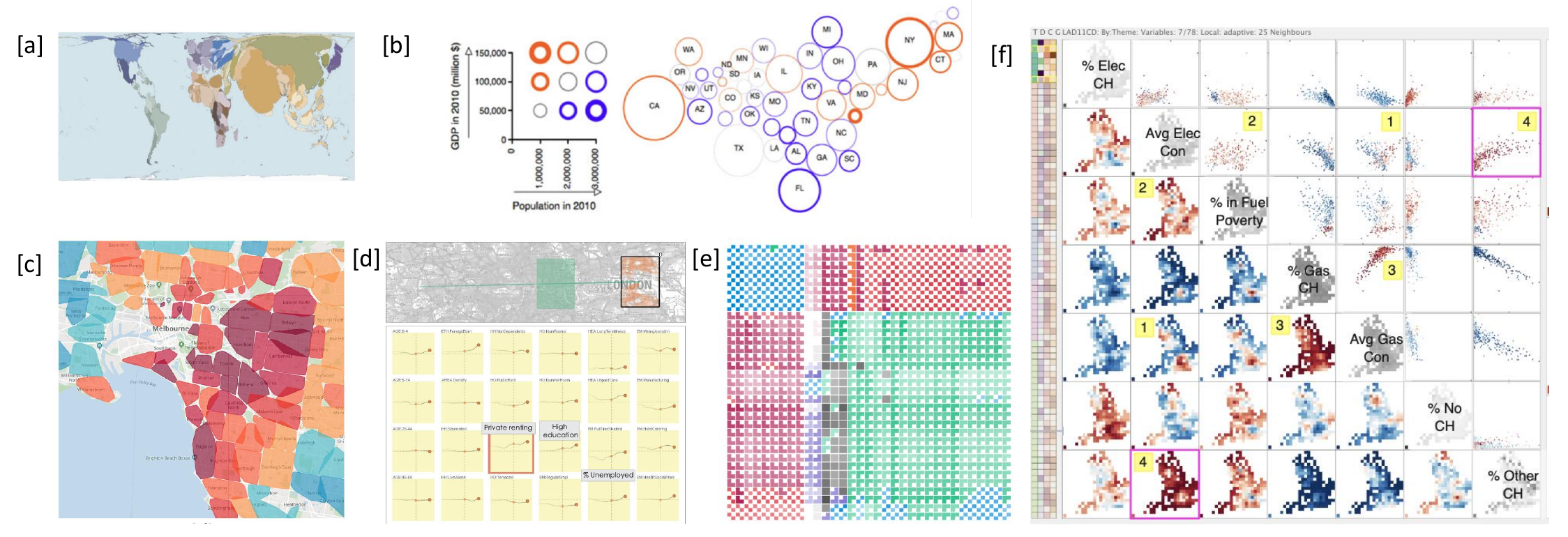}\\
    \caption{Visualizations for demographic information analysis with the spatial distribution (\textbf{a}: WORLDMAPPER~\cite{dorling2006worldmapper}, \textbf{b}: Nusrat et al.~\cite{nusrat2017cartogram}, \textbf{c}: ConcaveCubes~\cite{li2018concavecubes}, \textbf{d}: Attribute Signatures~\cite{turkay2014attribute}, \textbf{e}: Yates et al.~\cite{yates2014visualizing}, \textbf{f}:Goodwin et al.~\cite{goodwin2015visualizing}).
    }
    \label{fig_demographic}
    \end{center}
    \vspace{-.5cm}
\end{figure*}

\textbf{River Metaphor Visualization.}
Early on, TextFlow~\cite{cui2011textflow} and EvoRiver~\cite{sun2014evoriver} are introduced to track topic evolution such as political or journalistic subjects using a river metaphor visualization. However, topics could be subdivided into subtopics or keywords, and tracking this hierarchical topic evolution (e.g., split and merge) is challenging. Cui et al.~\cite{cui2014hierarchical} present RoseRiver, an interactive visual topic analysis approach to progressively explore complex evolving patterns of hierarchical themes. Fig.~\ref{fig_conversation}d is the RoseRiver result for the Prism scandal (June 5 to August 16, 2013). From the expert feedback, one political science professor evaluates that topic split and merge patterns allow him to track interaction between different topics, and it could be useful in public administration research.

\textbf{Glyph.} El-Assady et al.~\cite{el2016contovi} introduce ConToVi, a multi-party conversation exploration tool using glyphs and animation effects. 
It is a circular shape visualization that can track the movements of entities through animation within a conversation topic, called Topic-Space. Fig.~\ref{fig_conversation}e is its example case for Obama vs. Romney presidential debates in 2012. Each dot in Topic-Space represents a speaker's utterance and its colors indicate different speakers in the conversation. They point out that users might not be able to fully track the moving dots due to motion blindness. To overcome this limitation, they draw trail trajectories together following dot movements. Users could track their movement direction by reading the trails’ color opacity and width.

\begin{figure}[t!]
\begin{center}
    \includegraphics[width=1.0\columnwidth]{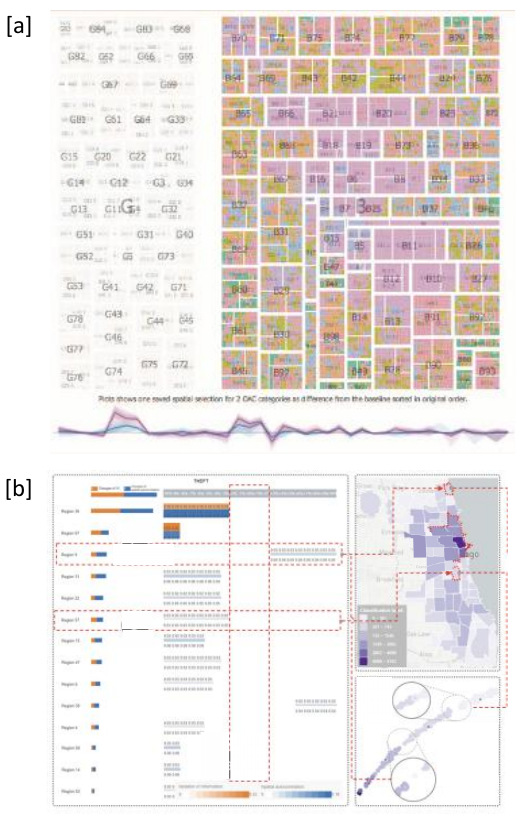}\\
    \caption{Visualization approaches to investigate the effect of uncertainties on political scientists' geo-demographic analysis. Uncertainties occur when a large number of demographic attributes are dimensionally reduced to manageable numbers (\textbf{a}: Slinsby et al.~\cite{slingsby2011exploring}, \textbf{b}: Huang et al.~\cite{huang2019exploring}).
    }
    \label{fig_demographic_uncertainty}
    \end{center}
    \vspace{-.5cm}
\end{figure}

\subsection{Demographic Information Analysis (A4)}
Understanding demographic information with spatial distribution allows political scientists to analyze dynamic links between politics, birth and mortality rates, and migration, taking into account multiple population characteristics, including age, gender, and religious composition~\cite{draper2008votes}.

\textbf{Map Visualization.} Map visualization is widely used due to its intuitiveness and efficiency in reporting values alongside geographic information. Dorling et al.~\cite{dorling2006worldmapper} first introduce WORLDMAPPER, a cartogram (Fig.~\ref{fig_demographic}a) in which the geometry of regions is distorted in order to convey the information of an attribute.
It has a limitation in that only one statistical variable per region could be encoded. In practice, however, it is often useful to display both variables simultaneously (e.g., Democrats vs. Republicans and GDP vs. Population). To this challenge, Nusrat et al.~\cite{nusrat2017cartogram} introduce a cartogram-style visualization allowing users to compare bivariate variables. Their approaches are shown in Fig.~\ref{fig_demographic}b. The ring colors and thickness are determined by the binary relationship between the two variables. \HDY{On one hand, Large red rings in Fig.~\ref{fig_demographic}b have a high population and a large GDP, with GDP dominating the comparison. On the other hand, large blue rings implies a large population and GDP, but population dominates the comparison.}

Li et al.~\cite{li2018concavecubes} point out that the earlier methods do not capture the real-world geography semantics (e.g., country, state, city), and proposed ConcaveCubes (Fig.~\ref{fig_demographic}c). It is a concave hull construction method that supports boundary-based cluster map visualization and maintains the informational integrity of real-world spatial semantics.

Many novel map visualizations have been introduced so far. However, visualization practitioners must choose map styles carefully. For example, PDViz in Fig.~\ref{fig_pdviz} chose a hex-bin map to display geographic proximity between the U.S. states. They choose it due to the possibility that state sizes could cause unexpected bias when examining the states and their corresponding properties.
The benefits and drawbacks of various map view styles have been discussed in numerous earlier papers~\cite{heilmann2004recmap, cano2015mosaic, hografer2020state}. 


\begin{figure*}[t]
\begin{center}
    \includegraphics[width=.9\textwidth]{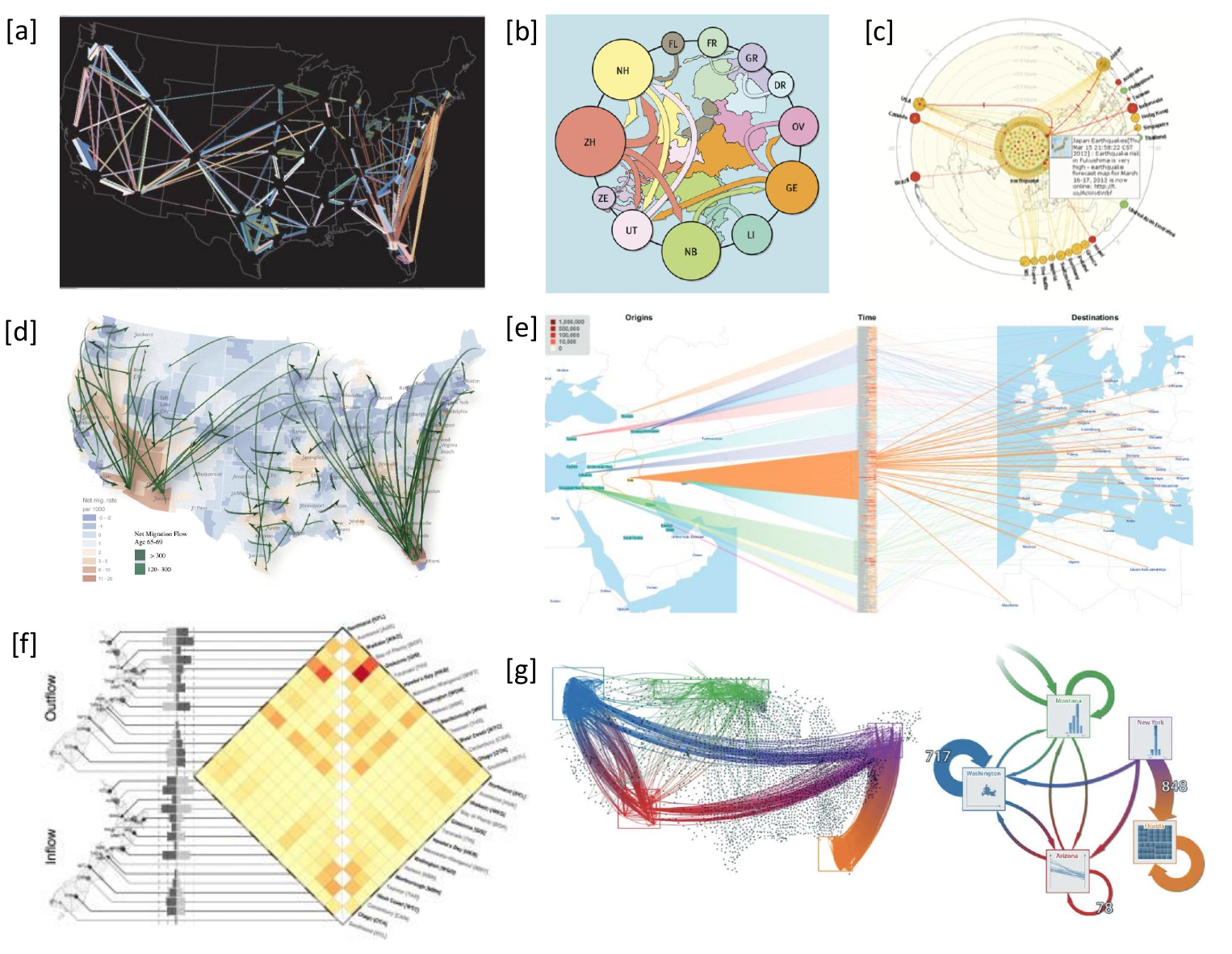}\\
    \caption{Visualizations for geographic information diffusion analysis (\textbf{a}: Guo et al.~\cite{guo2009flow}, \textbf{b}: Necklace Maps~\cite{speckmann2010necklace}, \textbf{c}: Whisper~\cite{cao2012whisper}, \textbf{d}: Guo and Zhu~\cite{guo2014origin}, \textbf{e}: Flowstrates~\cite{boyandin2011flowstrates}, \textbf{f}: MapTrix~\cite{yang2016many}, \textbf{g}: Elzen and Wijk~\cite{van2014multivariate}).
    }
   
    \label{fig_diffusion}
    \end{center}
    \vspace{-.5cm}
\end{figure*}

\textbf{Hybrid Approach.} It is challenging to analyze the multivariate geographical data by location. Turkay et al.~\cite{turkay2014attribute} introduce a visualization method, called Attribute Signatures (Fig.~\ref{fig_demographic}d). 
It consists of a geographical map and small multiples of line or bar charts. 
Users can determine the geographic scope to analyze by setting a path in the map view. 
For each element of small multiples, the x-axis represents the positions along the path on the map, and the y-axis shows the property values. Goodwin et al.~\cite{goodwin2015visualizing} and Yates et al.~\cite{yates2014visualizing} introduce visualizations integrating map and glyphs in the matrix form respectively (Fig.~\ref{fig_demographic}f and e). They enable users to gain a thorough understanding of local correlations and multivariate relationships.

Too much geo-demographic information in data would make analysis difficult to compare population characteristics by geographical areas.
As such, analysts sometimes use dimensionality reduction techniques in order to get a manageable set of variables. However, it inevitably hides the heterogeneity that changes within the demographic and geographic perspectives. 
\HDY{To tackle this challenge,} Slinsby et al.~\cite{slingsby2011exploring} introduce a visualization system consisting of a hierarchical matrix and a parallel coordinate plot as shown in Fig.~\ref{fig_demographic_uncertainty}a. It reveals these uncertainties for users' explainable analysis. Huang et al.~\cite{huang2019exploring} also propose a visualization system (Fig.~\ref{fig_demographic_uncertainty}b) with the same purpose but by using two maps and one scatter plot. \HDY{They demonstrate its usability using Chicago crime data~\cite{ChicagoDataPortal} to illustrate how attribute values for Chicago counties affect crime rates and how their uncertainty reshapes choropleth maps.} They report that domain experts in their study could have a better understanding of dimensionality reduction results and their uncertainties with the proposed systems.


\subsection{Geographic Information Diffusion Analysis (A5)}

Understanding the potential and dynamic interdependencies between political actors is one key subject in the field of political science. It includes an understanding of political public opinion and policy diffusion across spatial semantics.

\textbf{FlowMap Visualization.} A FlowMap visualization is useful to provide information on origin-destination (OD) regions and their associated features together by integrating map and network visualizations~\cite{cox19963d, munzner1996visualizing}. 
Tobler~\cite{tobler1987experiments} introduces Flow Mapper plotting volume scale bands between countries to express the directional connections and their quantitative features. Then, Guo et al.~\cite{guo2009flow} extend it to visualization of multivariate attribute flows by using multiple colors (Fig.~\ref{fig_diffusion}a). Flow maps~\cite{phan2005flow, buchin2011flow} have evolved in the direction of easing the clutter and representation of hierarchical structures of flows. 
Necklace Maps~\cite{speckmann2010necklace} and Guo and Zhu~\cite{guo2014origin} extend FlowMaps to explore multivariate data. They are shown in Fig.~\ref{fig_diffusion}b and Fig.~\ref{fig_diffusion}d. \HDY{Necklace Maps also facilitates users to easily compare the sizes of attributes (e.g., populations, energy imports, and estimated number of illegal immigrants)along the necklace in the map. } More details on the design principles of flow maps could be found in ~\cite{jenny2018design}.

Cao et al.~\cite{cao2012whisper} introduce Whisper, a different style of visualization using map and network together. It follows a sunflower metaphor as Fig.~\ref{fig_diffusion}c. It visualizes diffusions of temporal trends, socio-spatial coverage, and community responses in social media. 
\HDY{It allows users to examine who the media and opinion leaders are driving social and political reactions. They evaluate Whisper by interviewing with political scientists. } It is reported that Whisper effectively helps political scientists majoring in social networks and social influences to understand and track reactions to large-scale political events such as political campaigns and elections. 

\textbf{Hybrid Approach.} Another visual representation of geographic networks is an integration of a matrix and map. Wood et al.~\cite{wood2010visualisation} introduce its concept, and it mitigates the clutter problem of traditional flow maps. Boyandin et al. and Yang et al. extend OD map proposing Flowstrates~\cite{boyandin2011flowstrates} and MapTrix~\cite{yang2016many} respectively (Fig.~\ref{fig_diffusion}e and f). Both techniques contain two maps and one separate heatmap. The heatmap shows magnitudes of flowing attributes from source to destination. 
\HDY{The heatmap in Boyandin et al.'s approach additionally presents the temporal context, while MapTrix allows users to analyze multiple attributes' flow but one timestamp at a time.}
Despite the fact that Bundled Flow Map~\cite{pupyrev2012edge} and OD Map~\cite{wood2010visualisation} approaches are comparable to MapTrix, Yang et al. reported that MapTrix outperforms them in terms of visual clutter and scalability.  

Elzen and Wijk~\cite{van2014multivariate} introduce a visualization technique presenting high-level and infographic-style overviews. It is a hybrid visualization method of a network and one of the scatterplots, parallel coordinate plots, histograms, and treemaps. As shown in Fig.~\ref{fig_diffusion}g, it allows users to view outliers, patterns, and trends for combined elements. \HDY{For example, it can be applied to United States migration data~\cite{IRS}, and users can obtain an overview of incoming and outgoing migration movements between states and counties. }
Their approach has the advantage of making it simple to communicate analysis results to a bigger audience.

\section{Visualization Applications and Systems}
\label{sec_review_vis_system}

In this section, we review visualization applications designed for political scientists and analyze their system requirements. 
\HDY{In our data collection, 9 paper works introduce visualization applications in five domain interests, and 1 work visually supports query construction. This section is objective to provide an overview of domain interests and applications while summarizing task requirements. }
Understanding task requirements for target users is one of the important processes to fully support their tasks and to implement better visualization systems. System requirements usually are discussed in collaboration with relevant stakeholders in design processes~\cite{sacha2014knowledge}. 



\subsection{Public Policy Diffusion}

\begin{figure}[t]
\begin{center}
    \includegraphics[width=\columnwidth]{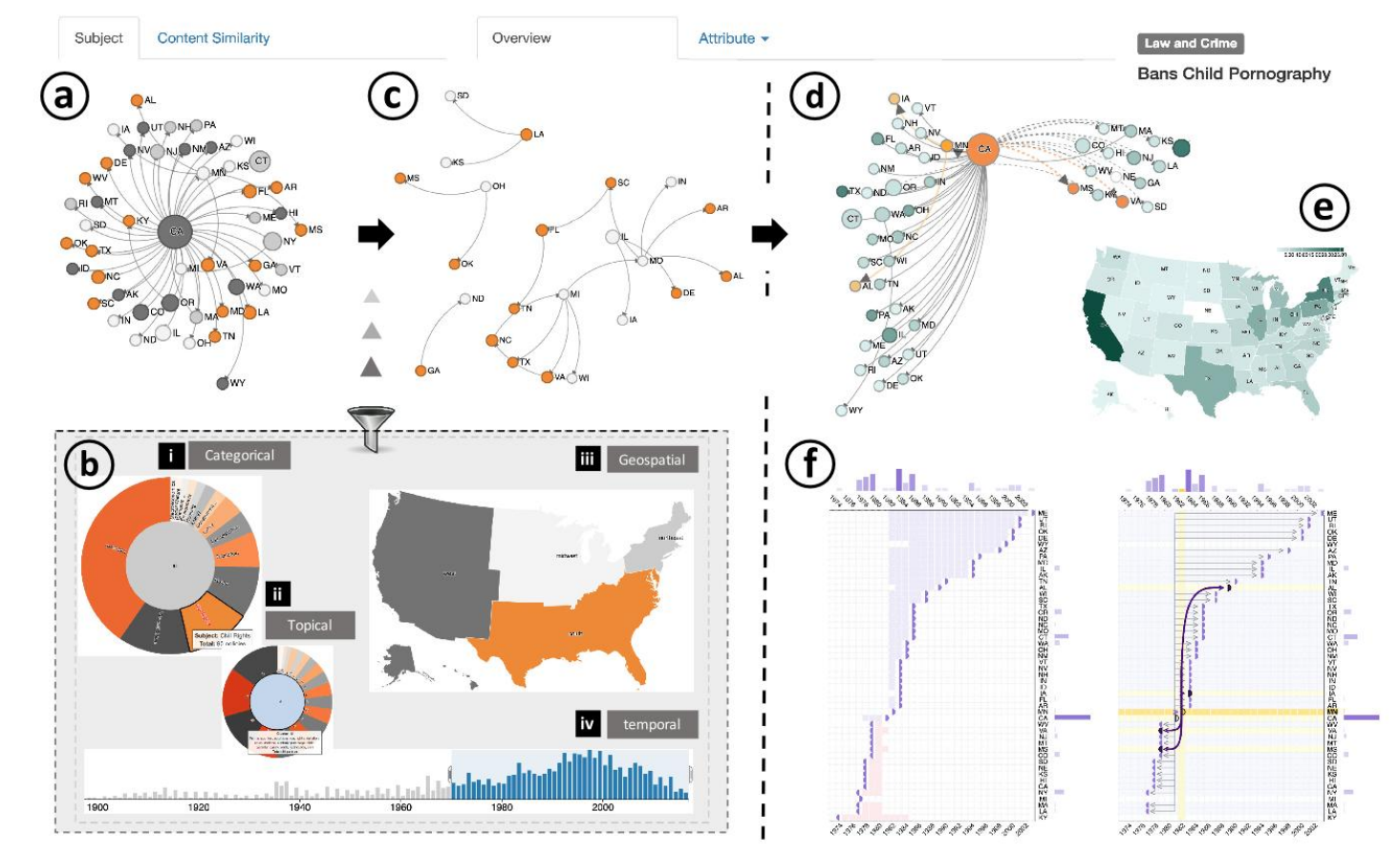}\\
    \caption{PolicyFlow~\cite{ahn2020policyflow} allows users to analyze the policy diffusion patterns across geospatial and temporal contexts as well as political contexts like policy topics.
    }
    \label{fig_poliyflow}
    \end{center}
\end{figure}

\begin{figure*}[t]
\begin{center}
    \includegraphics[width=.93\textwidth]{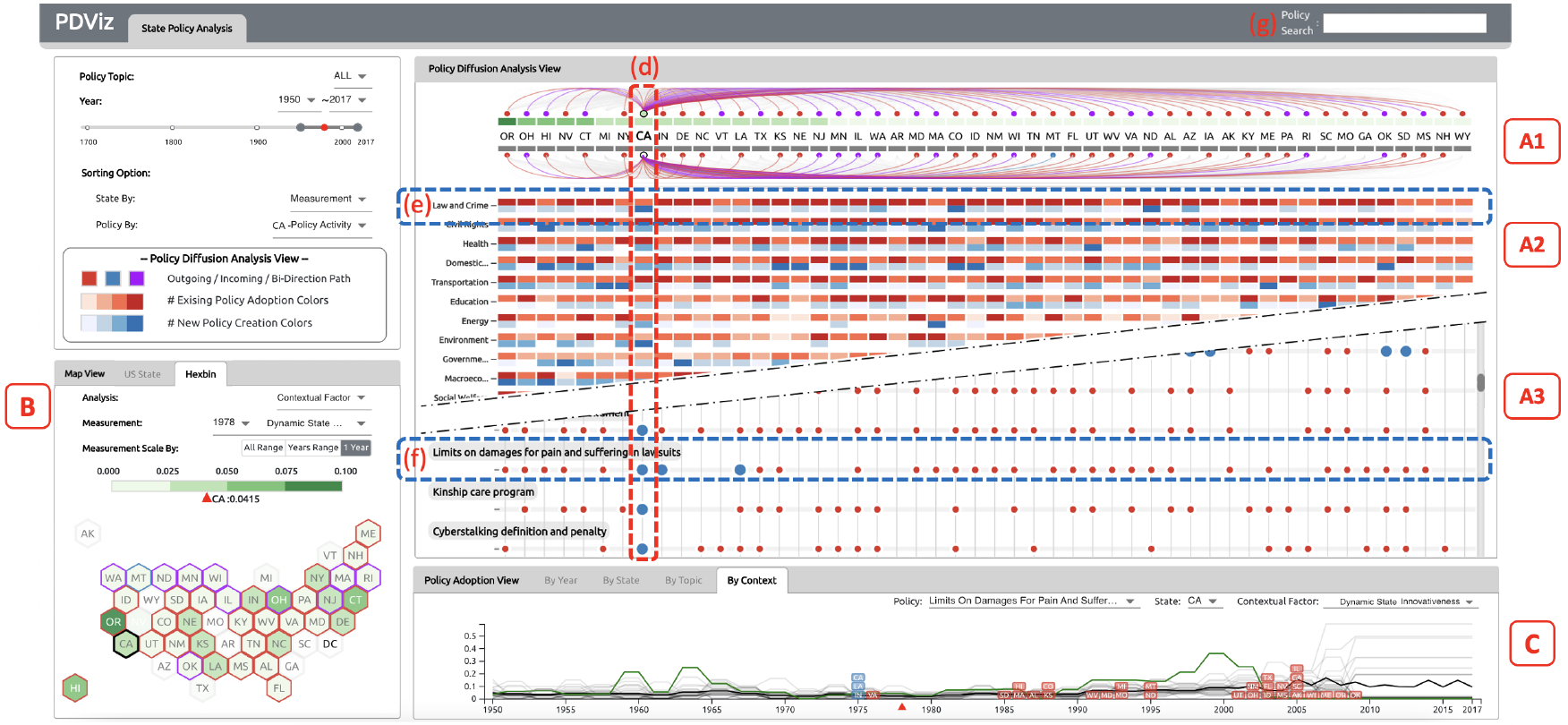}
    \caption{PDViz~\cite{han2023} is a visual analytics approach for political scientists to dynamically analyze the policy diffusion history and underlying patterns.
    }
    \label{fig_pdviz}
    \end{center}
\end{figure*}

Public policies affect everyone's daily lives within national and state borders in terms of welfare, education, economy, and safety. For instance, it oversees the location of new businesses, the conduct of elections, and the decision to raise the minimum wage. Different nations and states have adopted different policies to provide public services and address community issues due to their diverse geographical and demographic characteristics. Under these diversities, it is critical to understand how the policies have been spread across the nations and states because they repeatedly face similar political situations and uncertainties. Understanding the potential and dynamic interdependencies between society members is one of the major topics in the political science field.

PolicyFlow~\cite{ahn2020policyflow} and PDViz~\cite{han2023} are two visualization applications that have been introduced to support policy diffusion analysis. According to the two works, political scientists' tasks in the analysis are the followings: \textbf{T1. (Retrieve Value)} Political scientists identify policy spread patterns and innovative nations/states which have created new policies. As policies spread over time, there are many connections between states, resulting in a complex network. They are interested in answering questions like ``Which state is the source of policy diffusion to other states?'', ``What are the statistics on policy adoption by year, by state, by subject?'', and ``Are some states in the US more innovative than others?''; \textbf{T2. (Compute Derived Value)} They explore and identify policy diffusion in a geographic and temporal context. Specific questions are ``Do innovative states cluster into regions of innovation?'' and ``How are impacts of the states in the policy diffusion examined over time?''; \textbf{T3. (Filter, Characterize Distribution)} They compare policy diffusion in thematic contexts. They are interested in identifying which states were policy leaders (or laggards) in different policy topics and how each state responded to the new wave of policies being adopted. Example questions are ``How do policy leaders differ across policy topics?'' and ``How do diffusion patterns differ across policy topics?'' \textbf{T4. (Correlate)} Political scientists also analyze the diffusion in sociopolitical factors including foreign-born and crime rates.

\begin{table}[t!]
\begin{center}
    \includegraphics[width=.5\textwidth]{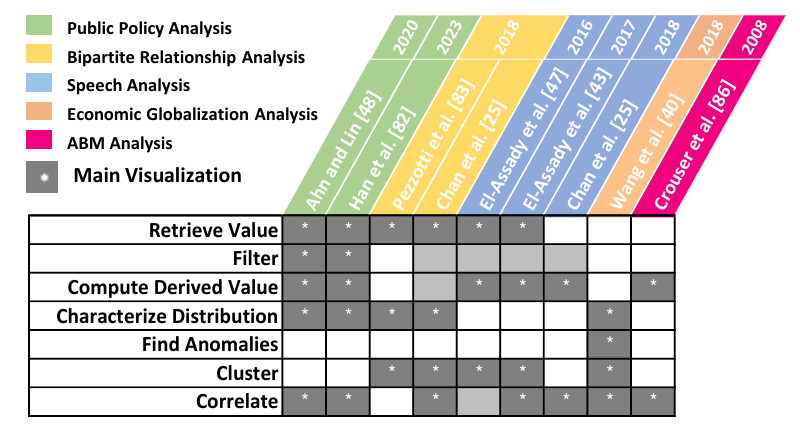}\\
    \caption{9 visualization applications are studied for political scientists in 5 different domains. Darker gray cells with a white star indicate the major tasks which the application support. Lighter cell tasks are also supported in the application, but not major features.
    }   
    \label{fig_poliyDiffusion}
    \end{center}
\end{table}

\subsection{Bipartite Relationship Analysis}

A bipartite graph is a powerful method to model relationships between two groups~\cite{pezzotti2018multiscale}. Political scientists, for example, model key relationships from bipartite graphs of real-world data, such as legislators' votes on legislation and individuals’ affiliations in various social groups. Visualizing bipartite graphs allows users to understand the interrelationships between the elements of two social/political groups. It could also be used for identifying clusters of similarly connected elements.

VIBR~\cite{chan2018v} and Pezzotti et al.~\cite{pezzotti2018multiscale} report the following political scientists’ tasks in bipartite graph analysis. \textbf{T1. (Characterize Distribution, Cluster)} They identify clusters of similar elements in one group with respect to connections to elements in another group and vice versa. For instance, recognizing bills on which similar members voted and identifying lawmakers who voted similarly on a particular bill are included. Next, \textbf{T2. (Retrieve Value, Correlate)} the political scientists analyze the interrelations between clusters in two collections. They look at the properties of the nodes to determine how they relate to one another in a bipartite connection.

\subsection{Speech Analysis}
Central to the understanding of political representation is a stable understanding and measurement of public opinion. Multilateral conversations, such as political debates, oral court arguments, and public opinions in social networks, are characterized by rapid exchange of opinions and information. The conversations produce long verbatim textual records, which are full of interruptions, inflections, repetitions, etc. Deliberative democracy relies on these discussions and conversations to reach mutual consent. 

Visualization approaches to assist with the conversation/opinion analysis has been widely studied in the early 2010s~\cite{wan2014improving}. Among them, we found three visualization approaches particularly designed with and for political scientists, VIBR~\cite{chan2018v}, ConToVi~\cite{el2016contovi}, and NEREx~\cite{el2017nerex}. They reported following political scientists' tasks. \textbf{T1.} Political scientists first try to create transcripts or understand the content of the conversations. This is usually done by watching a video or listening to an audio recording of a discussion. This task allows them to catch up with an overview of the context and dynamics of the conversation. \textbf{T2. (Retrieve Value, Filter, Cluster)} From the overview, they identify specific topics of interest for their analytic work. \textbf{T3. (Correlate)} They analyze the dynamics of conversation of single or multiple speakers and examine speaker interactions across discussion topics. This includes identifying speakers’ emotional context of entities and highlighting politeness. \textbf{T4. (Compute Derived Value)} They also analyze each speaker's contribution or impact on each topic over time.

\subsection{Economic Globalization and Stability}
\begin{figure*}[t!]
    \centering
    \includegraphics[width=0.92\textwidth]{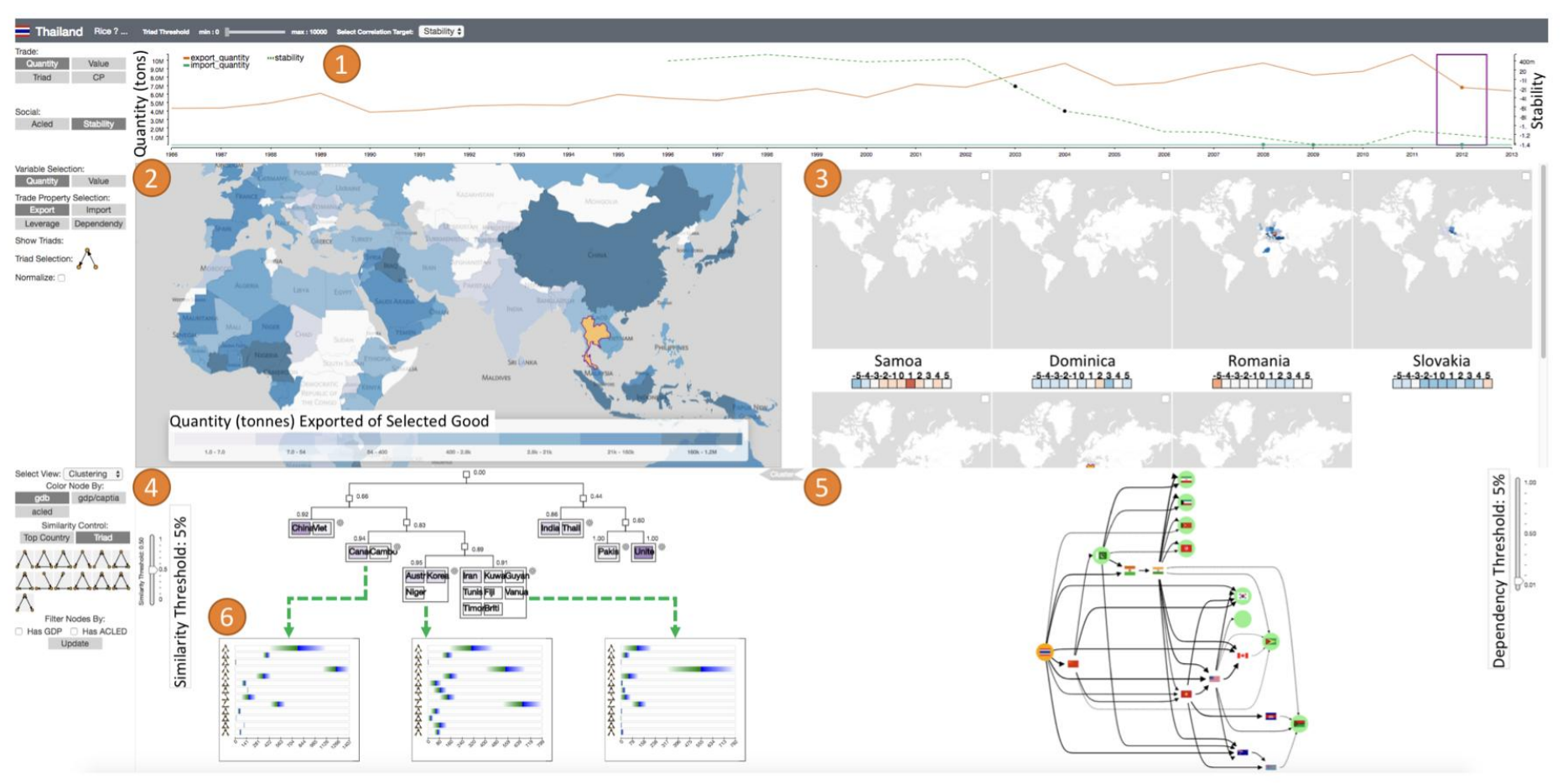}
    \caption{ A visual analytics framework for exploring global trade networks. It allows users to explore global trade relationships and to identify temporal anomalies within the trade network~\cite{wang2018visual}.
    }
    \label{fig_global_trade}
\end{figure*}

Economic globalization increases global connectivity while also creating complex interdependencies among various supply chains. Economic globalization refers to the international trade of products, capital, services, technology, and information among countries. Researchers in political science and international affairs investigate the complex dynamics of global trade networks and how their structure relates to regional instability. According to earlier research~\cite{hsiang2011civil, marshall2008fragility}, abrupt changes in these dependency networks could be utilized as an indicator of potential risks of violence and armed conflict. Political scientists examine international relationships to answer the question, ``Are there topological network structures associated with a trade that serve as potential indicators of future instability?'' However, understanding those complex dynamics and network structures is a challenging task. 

Wang et al.~\cite{wang2018visual} propose a visualization interface to ease this challenge by collaborating with political scientists. They report political scientists' two main tasks. \textbf{T1. (Find Anomalies, Cluster)} They investigate complex dynamics within global trade networks and analyze how these network structures relate to regional instability. They specifically conduct a triad analysis of trade patterns, which is a typical method in the study of trade networks. A triad is a three-node directed subgraph of the overall network. Each triad can be regarded as a group of three countries and related trade interactions. The experts perform anomaly detection, correlation analysis, clustering, and similarity comparison within the networks. \textbf{T2. (Retrieve Value, Correlate)} Based on the triad analysis method, they analyze dependencies and leverage between countries. They also examine the volume and proportion of trade events for imports and exports between the countries in a triad.  It allows them to analyze trade network properties such as structural balance, transitivity, and vulnerability.

\begin{figure}[t]
\begin{center}
    \includegraphics[width=.5\textwidth]{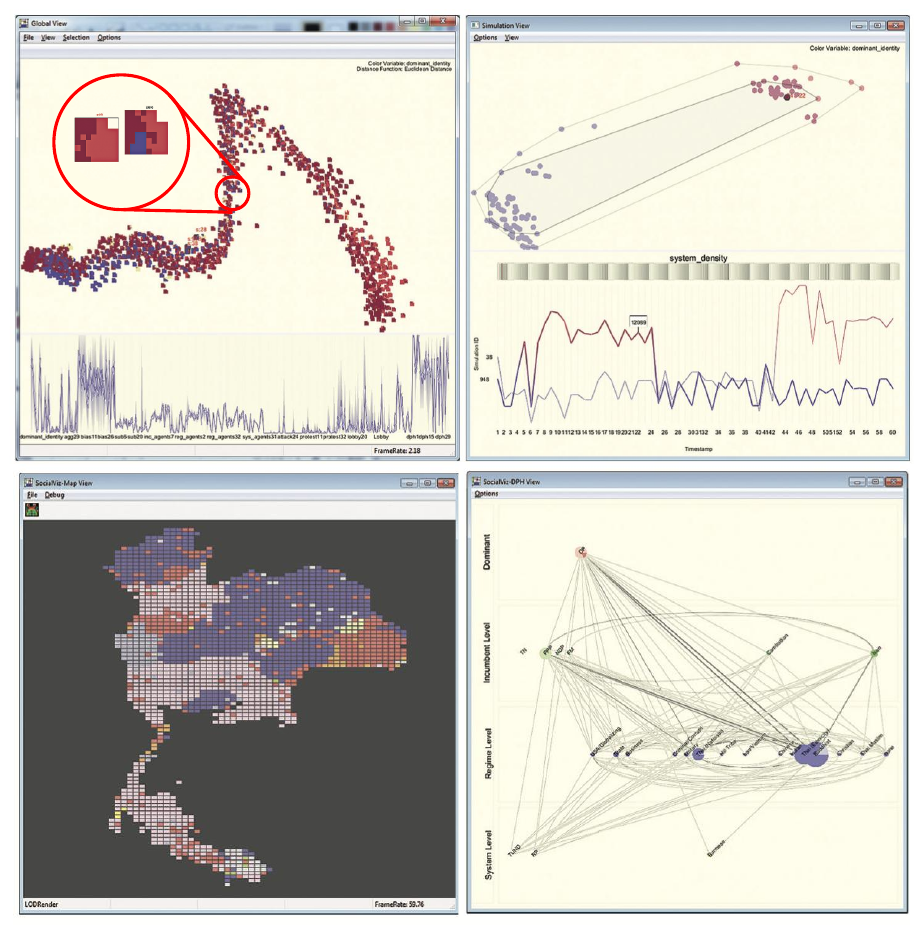}\\
    
    \caption{SocialVis and MDSVis~\cite{crouser2011two}, visual analytics approaches to interpreting agent-based modeling (ABM) results. They could support political scientists to identify a country’s political patterns, which might indicate the imminence of civil unrest and help predict catastrophic events.
    }
    \label{fig_poliyDiffusion}
    \end{center}
\end{figure}

\subsection{Agent-Based Simulation Result Analysis}

Political science sometimes uses agent-based modeling (ABM) to simulate and investigate social behaviors between society members. An ABM simulation is a computational model of autonomous entities or agents that interact with each other according to a set of rules and goals according to a behavioral system based on complex mathematical and statistical specifications. Political scientists, for instance, use these simulation models to examine the likelihood of civil unrest caused by political situations in terms of social behaviors such as collaboration, conflict, and violence.

According to Crouser et al.~\cite{crouser2011two}, political scientists have two major tasks. \textbf{T1. (Compute Derived Value)} Political scientists aim to develop or evaluate political theories with ABM simulations. They feed potential behavioral patterns as numerical parameters into the models based on observed real-world behaviors and demographic information. \textbf{T2. (Correlate)} They examine the model outputs to find coherent narratives that describe relevant interactions and identify intriguing or likely hypotheses. This entails investigating and contrasting the outcomes of various scenarios, as well as incorporating domain knowledge into data analysis to produce initial hypotheses for both single- and multiple-simulation scenarios.

\subsection{Other Visual Analytics Support Tool}

Visual analytics systems could also help users' data scheme structure processes from multiple data sources. Cashman et al.~\cite{cashman2020cava} present CAVA, a visual analytics system to explore properties visually and to serve as an interface for constructing queries. They evaluate their approach by presenting a usage scenario for political scientists who are interested in armed conflict events. \HDY{
From a given dataset, political scientists are intereseted in which attributes could be relevant. For example, one could speculate that economic indicators would influence the sorts and severity of conflict events. CAVA visually supports him or her to implement queries exploring correlation between the attributes and generates various visualizations.
}
They also conduct a user study that confirm the effectiveness of CAVA in performing data augmentation as part of a visual analytics pipeline.




\subsection{Visual Analytic Task Requirements}
\HDY{We summarize the task requirements of political scientists for visualization researchers and practioners in visual analytics terms.}

\textbf{Visualized Overview.} The motivation behind the overview is to get a quick exploration of data. It is essential to provide an overview because the size of the data to be analyzed is big. Political science typically uses collected data for analysis rather than real-time data. Visualization for real-time data would not necessary to be taken into account. A visualized overview can provide a starting point for political scientists to begin their analysis. It could allow them to find extreme values (e.g., anomalies, priorities, outliers, etc) with visual cues such as shapes, colors, and positions. Overviews could also help them identify patterns.


\begin{table*}[t!]
    \scriptsize
    \begin{center}
    \begin{tabular}{|p{4.4cm}|p{1cm}  |p{2.5cm}  |p{7.8cm}    |}
        \hline
        \textbf{Visualization Type} & \textbf{Count} & \textbf{Usage Percentage} & \textbf{Conveying Knowledge (Count)} \\ 
        \hline
        Dot Chart & 305 & 37.2\% & Association (264), Value/Derived Value (20)\\
        \hline
        Line Chart & 176 & 21.5\% & Association (52), Trend (42), Value/Derived Value (33)\\
        \hline
        Scatter Plot & 96 & 11.7\% & Association (47), Trend (18), Value/Derived Value (12)\\
        \hline
        Bar Chart & 82 & 10.0\% & Difference (32), Value/Derived Value (32)\\
        \hline
        Histogram \& Density Chart & 56 & 6.83\% & Distribution (56)\\
        \hline
        Choropleth Map & 33 & 4.02\% &  Value/Derived Value (33)\\
        \hline
        (100\%) Stacked Bar Chart & 6 & .97\% & Difference (3), Value/Derived Value (3)\\
        \hline
        (Stacked) Area Chart & 4 & .49\% & Difference (2), Value/Derived Value (2),  \\
        \hline
        Box Plot & 4 & .49\% & Association (2), Distribution (2)\\
        \hline
        Bubble Chart & 1 & .12\% & Value/Derived Value (1)\\
        \hline
        Pie Chart & - & - & -\\
        \hline
        Cartogram & - & - & -\\
        \hline
        Treemap & - & - & - \\
        \hline
        Other (Network, Gantt chart, Wordcloud, Matrix, Contour and Point map) & 19 & 2.31\% & Value/Derived Value (8),  Relationship (4), Category (2), Association (2), Cluster, (1), Location (1), Range (1) \\
        \hline
        \end{tabular}

    \label{tab:vis_type_taxonomy}
    \end{center}
      \caption{Visualization types~\cite{lee2016vlat, traunmuller2020visualizing} and their usage count, percentage, and knowledge types from the three political science journals }
\end{table*}

\textbf{Dynamic Data Exploration.} It includes a variety of analytical tasks: filtering, sorting, characterizing distributions, and retrieving values. Filtering and sorting allow users to narrow down the range of data from an entire dataset to suit their interests. It also allows political scientists to explore data in different contexts such as geographic, temporal, thematic contexts, and even sociopolitical factors. The traditional method for political scientists is using a query-based approach from a database. This requires a time-consuming workload to create and validate accurate queries. In contrast, visualization systems could support easy and intuitive data exploration~\cite{cashman2020cava}. This could be achieved easily by visualizations or via HTML elements like drop-down menus and sliders. For example, VIBR provides a drill-down function that allows users to see more detailed content by clicking on a given visualization element in the overview. It also provides HTML elements to adjust the density of visualized overview and cluster size for a deeper exploration. PDViz, for another example, helps political scientists identify different policy diffusion patterns by states and by policy topics to analyze the impacts of the sociopolitical factors on states’ policy adoption.

In technical-wise, political scientists want a function to upload datasets and download analyzed results in visualization systems.

\textbf{Comparative Analysis.} Political scientists are interested in finding and comparing relationships between political actors. The comparative analysis includes cluster and correlation analysis. Cluster analysis helps them identify clusters of entities having similar properties. Correlation analysis finds useful relationships, associations, and patterns between two or more entities from given data cases. The cluster and correlation analysis are essential for political scientists to establish or evaluate political hypotheses based on observed real-world data. Those analyses highly depend on computational algorithms. For example, Wang et al.~\cite{wang2018visual} adopted a triad pattern analysis method widely used in global trade network analysis. Triad patterns enable political scientists to pinpoint nations with comparable trade patterns and stability vulnerabilities in a given global trade network. 

\textbf{Set up a Hypothesis and Predictive Analysis.} Political scientists gain knowledge about patterns, events, and anomalies through comparative analysis based on historical data. From that knowledge, political scientists establish hypotheses or conduct predictive analysis. They usually judge the relationships based on statistical values which vary depending on the political science sub-domains. Visualization systems should provide statistical numerical values to help the users formulate hypotheses and do predictive analysis. This is a deriving value task. For example, PDViz~\cite{han2023} provides the Cox model\cite{jones2005beyond} results called hazard ratio along with visualizations to reveal how much a relationship is significant and credible between a state’s political adoption and sociopolitical factors. The hazard ratio indicates the relative impacts of the factors on the adoption probability.

\section{Perspectives on Data Visualization in Political Science}\label{sec_perspective_political_science}



In this section, we examine the political science perspective on data visualization in two different ways. We start by reviewing articles from journals in political science. Next, we look at the current state of data visualization education for aspiring political scientists.

\subsection{Visualizations in Political Science Publications}

We review articles published in what are regarded as the top three political science journals in 2021. The three journals are the American Political Science Review (APSR), the American Journal of Political Science (AJPS), and The Journal of Politics (JOP). These are the three most prominent general journals in political science, each being the flagship journal of its respective professional association: the American Political Science Association, the Midwest Political Science Association, and the Southern Political Science Association, respectively \cite{tyranny}. These are the three largest associations in the disciplines, and these journals are included in membership fees. They are extremely accessible to a wide range of political scientists. The APSR alone reaches 12,000 political scientists in 80 different countries. Furthermore, these journals do not discriminate against the type of publication based on methodological approaches, subfields of interest, or the time period being studied. They are broad, general journals that repeatedly have some of the highest impact factors scores for single-discipline journals in Political Science. Other journals in the field will limit publications to a specific subfield and sometimes methodological approach. 
This section reports which visualization types are familiar to political scientists and reviews the knowledge types that a visualization type provides.
We categorize them into one or more visualization and knowledge types as Table~\ref{tab:vis_type_taxonomy} and ~\ref{tab:knowledge_type_taxonomy}. 
The total number of papers we review is 264, and we found 820 unique visualizations. 


\begin{table}
    \scriptsize
    \begin{center}
    \begin{tabular}{|p{1.5cm}|p{6.2cm}|}
        \hline
        \textbf{Knowledge Type} & \textbf{Example} \\ 
        \hline
        (Derived) Value &  The average salary of graduated students in laws school is 60k per year \\
        \hline
        Distribution &  The distribution of consumption month by month in Italy is fairly even \\
        \hline
        Difference &  In USC, there is still a greater absolute enrollment in the social sciences than the biological sciences\\
        \hline
        Categories &  All in all, jobs in this data can be classified into 4 categories: rich, middle, lower middle, and lower\\
        \hline
        Cluster &  Countries in Western Europe tend to group together according to their consumption amounts in 1999 \\
        \hline
        Association &  In the US, there is a negative correlation between income and obesity when income is less than 50k \\
        \hline
        Trend &  Veterans' benefits are going down over the past ten years\\
        \hline
        \end{tabular}
    \label{tab:knowledge_type_taxonomy}
    \end{center}
    \caption{7 Knowledge type taxonomies from Chen et al.~\cite{chen2009toward}}
\end{table}

\subsubsection{Categorization}\label{sec_categories}

We categorize visualizations into knowledge types. 
The taxonomies for visualizations and knowledge types and their short descriptions are in Table~\ref{tab:vis_type_taxonomy} and ~\ref{tab:knowledge_type_taxonomy}. The visualization types are categorized as one of 13 types and `Other.' The 13 visualization types are from the combination of explored types in visualization publications and Traunmüller's work~\cite{lee2016vlat, traunmuller2020visualizing}. 'Other' includes charts that illustrate theoretical or mathematical models and diagrams that are not categorized into 13 types. To categorize visualizations by knowledge types, 7 knowledge types are used. It is originally defined by Chen et al.~\cite{chen2009toward}.


\subsubsection{Results}

The visualizations of dot charts, line charts, scatter plots, and bar charts are the most used graphical format in political science. The summary is presented in Table~\ref{tab:vis_type_taxonomy}. First of all, the dot chart charges about 37.2\% of all graph usage.
It is used to report the type of knowledge of association, such as degrees of coefficient or marginal effects. The second most popular format is the line chart (21.5\%). It conveys the authors' findings related to association, trend, and (derived) values. The scatter plot (11.7\%) and the bar chart (10.0\%) are followed. 
In particular, when dot charts and scatter plots are used for association purposes, they have additional lines of best fit to present the estimated effects of variables and the associated uncertainties. The bar graph is used to present the differences between variables, and a histogram is used for the distribution of the data. Among the bar charts, some of them are represented in 3D. 
We find 33 choropleth map usages, but no cartogram map is found. Other formats include the stacked bar chart, (stacked) area chart, box plot, and bubble chart which were used less than 1\% each. 


Our observations find that about 32.3\% of the visualizations are used individually. Among the single plots, some of them are represented in the mixture design, including a scatter plot with regression lines, a scatter chart with a box plot, and a scatter plot or line chart with a rug plot. The rug plot is a subplot along the x- or y-axis that visualizes the distribution of data, analogous to a one-dimensional histogram or scatter plot. On the other hand, political scientists occasionally report results using multiple visualizations in a single figure (67.7\%). The combined forms include small multiple designs (60.76\%) aligning the same type of visualization, and multiple chart designs (6.94\%) aligning more than one type of visualization. 

Political scientists use textures, shapes, and color saturation to represent different categories of entities in visualizations. However, they highly depend on textures and shapes. This could be due to the traditional legacy in political science journals, where papers have been published in black and white \cite{traunmuller2020visualizing}. 
We find that only about 27.5\% of all figures used colors. Interestingly, some of the figures in single manuscripts are represented in colors, but the rest are not. This is most likely an indication of the journals charging a fee for color printing by figure.

\begin{figure*}[t!]
    \centering
    \includegraphics[width=\textwidth]{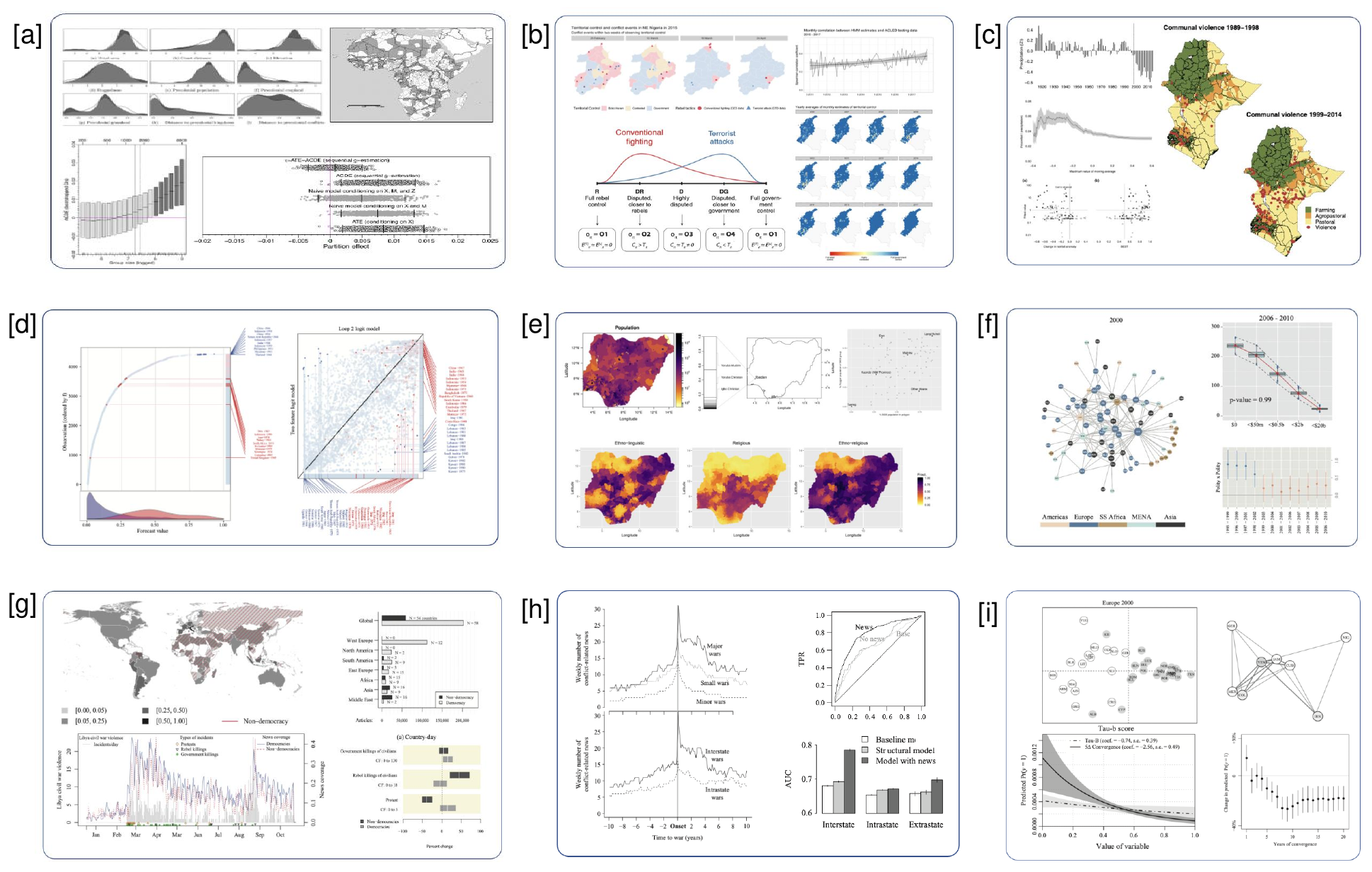}
    \caption{Figures from the papers that were awarded the JPR Best Visualization Award from 2013 to 2021 (\textbf{a-i}:~\cite{ito2021does, anders2020territorial, van2019climate, muller2018new, colaresi2017robot, kinne2016agreeing, baum2015filtering, chadefaux2014early, kinne2013igo}).
    }
    \label{fig:bestvis}
\end{figure*}

According to Traunmüller~\cite{traunmuller2020visualizing}, the usage of visualization in the field of political science is expanding once more these days.
He investigates the use of visualization in AJPS~\cite{AJPS} and finds an upward trend in its utilization between February 2003 and March 2018. He says that reporting results in visualization formats have increased to an average of three and a half in 2018 from one in 2003. 
In addition, The Journal of Peace Research (JPR) has been awarded the JPR Best Visualization Award \cite{visbest} since 2013 for papers (Figure ~\ref{fig:bestvis}) which explain procedures, data, and results effectively through graphical formats.


\subsection{Learning Visualization in Graduate Training in Political Science}

To explore how political scientists are trained in data visualization, we use the Political Science Rankings of the US News and Reports \cite{USNEWSPOL} to identify the top 20 ranked institutions in the US. We began by exploring the elements required in the research methodology curriculum and methods related to electives. This sample is an appropriate avenue to examine the training patterns and subsequent use of visualization in their published work. A 2012 study of all Ph.D. placements from political science departments found that the top 20 programs place 89.03\% of faculty members in the top 5 programs, 86.9\% in the top 20, and 62.28\% across all university types \cite{oprisko2012}.

From our analysis, only 6 programs explicitly state training in visualization or presentation as a core objective in one of their courses. Within these programs, the University of California at Berkeley (2), Washington University in St. Louis (3), and the University of California at San Diego (3) are the only programs with multiple courses that mention training in the presentation of results or data visualization. The other schools included in our audit are Duke University, the Ohio State University, and Columbia University with one course that meets our criteria. 

The visualization approach to these courses is also varied. Many of these approaches are directed graphs in the area of Game Theory and Formal Modeling. Other areas include causal pathway graphs, network graphs, and results from statistical analysis. Interestingly, of the six programs we included in this analysis, Washington, California, Duke, and Columbia are ranked in the top 10 of the U.S. News and Report rankings for Political Methodology \cite{USNEWSMET}, while Ohio State (13) and San Diego (17) just miss out. 

\section{Discussion}\label{sec_discussion}
\subsection{\HDY{Visualization Benefits for Political Scientists}}

Our review demonstrates that visualization systems introduced by visualization researchers could be beneficial to political scientists in the following stages~\cite{borkin2013makes}: (1) Exploring and familiarizing large data sets; (2) Identifying trends and unexpected patterns; (3) Generating hypotheses; and (4) Analyzing and verifying statistical findings. In addition, visualizations can lead to a potentially transparent analytic process for data. By demonstrating how datasets having the same statistical values could have various distributions, Matejka and Fitzmaurice~\cite{matejka2017same} contend that only providing statistical values could increase the risk of misinterpretation of the data. Additionally, data visualization has a superiority phenomenon that makes data more memorable than expressing it in tables or words~\cite{borkin2013makes}.

\HDY{The value of data visualization is supporting not only exploratory data analysis but also information communication and sharing.}
\HDY{The earlier sections primarily focus on data visualizations to support exploratory data analysis rather than to communicate information. Analysis and communication are two specific and unique purposes for data visualization.}
We also consider the importance of data visualization as a medium of communication between political scientists and other political actors, such as politicians and the general public, to be underestimated. The sharing of ideas and knowledge among political scientists can be facilitated by visualization. To uncover embedded insights from vast and complex data, collaborative efforts among domain experts including political scientists are required. Xu et al.~\cite{xu2018chart} and Li et al.~\cite{li2020resolving} demonstrated that visualization approaches could be advantageous for collaborators to review and analyze data stories created by other collaborative analysts. It could prevent them from repeating individual analysis which has already been completed and improve analytic efficiency by a meta-visualization approach to predict, cluster, filter, and connect all collaborators' outcomes. \HDY{
Other than the research perspective, there are practical infographic examples~\cite{ArcGISBlog, harrie2021thematic} for political scientists to share their ideas with other political actors. 
}

Finally, we also expect that visual analytics may have a role to play in the solicitation of data from research subjects. Providing interactive visualization tools to survey respondents, for instance, may help researchers to provide respondents with more information than they are otherwise able to by allowing those respondents to ``drive'' their information ingestion. In other words, visual analytics may prove a more appealing information transmission method for survey respondents who need to consume a large amount of data prior to answering a questionnaire than would the equivalent amount of text or static images. The ability to track the respondent as they interact with visual analytics tools may also prove helpful for political scientists interested in the saliency of different items about which information is available through the tool. 

\subsection{Knowledge Gap in Visualization}\label{sec_knowledgegap}
Political scientists are becoming more familiar with visual analytics tools, specifically, the increasingly common web-based dashboards associated with major data collection projects~\cite{ComparativeAgendasProject, VoteView, CSPP}. For instance, the Armed Conflict Location and Event Data Project (ACLED) maintains a dashboard for users interested in quickly assessing the scope of and trends present in their data \cite{raleigh:etal:2010}. The Peace Research Institute of Oslo maintains a similar dashboard associated with their GRID dataset \cite{tollefsen:etal:2012}. The Manifesto Project provides four ``simple dashboards'' on their website and links to external researchers' dashboards based on Manifesto data \cite{merz:etal:2016}. These tools, and others like them, allow researchers to inspect the available data quickly and assess whether it is suitable for their research purposes. \HDY{We find two major barriers between scholars in political science and data visualization.}

\textbf{\HDY{Political Scientists still prefer to use the table format.}}
We find that political scientists prefer to report numerical values in the table format and use visualizations mainly for reporting results instead of analyzing data. According to Kastellec and Leoni~\cite{kastellec2007using}, however, political scientists still prefer tables to graphs for the following reasons: (1) tables are easier to create than graphics; (2) tables are the standard format in political science for teaching and presenting results; (3) tables are useful to aid in replication studies because they convey exact numerical values. In addition, they noted that political scientists might have believed that utilizing graphical formats is inappropriate because they require more space than tables and are difficult to use when attempting to represent multiple regression results for multivariate data sets.


\textbf{\HDY{The discrepancy in trust in visualization approaches is exacerbated by a knowledge gap between visualization professionals and political scientists.}} We believe a knowledge gap between visualization experts and political scientists contributes to the tabular format serving as the primary method in the field of political science, not only the aforementioned three reasons for the use of the table formats. 
Dasgupta et al.~\cite{dasgupta2016familiarity} note that one barrier is the lack of trust in visualization approaches. 

Data is often seen as a collection of facts about the world, but deception in visualizations could happen at two levels: the chart level and the interpretation level. Depending on the intent of the people delivering the data, data visualization is subject to contamination in visual encoding processes. However, visualization researchers and practitioners aim to design tools to provide precise visual cues in order to support domain experts’ work and answer their scientific questions. For domain experts’ easy access to visualization representations, visualization practitioners often exclude data uncertainty or predictive variability~\cite{boukhelifa2009uncertainty, greis2017designing}. However, recognizing data uncertainty is crucial for data-based decision-making because it affects data credibility and judgment confidence~\cite{wesslen2021effect, sacha2015role}. This transparency gives researchers confidence when they want to know methodological or outcome details and promotes confidence in the visualization. The chart-level misunderstanding can be adjusted by collaborating with visualization researchers and political scientists.

The interpretation level is highly dependent on a user's visual literacy and familiarity with visualizations. According to the three political scientists among the authors, the most frequently used tools to create visualizations are ggplot2 \cite{ggplot}, R Shiny \cite{rShiney}, Stata \cite{Stata}, and Excel \cite{Excel}. However, many visualization tools (e.g., Tableau \cite{Tableau}, TIBCO Spotfire \cite{Spotfire}, and ADVIZOR \cite{ADVIZOR}) have emerged these days, allowing them to construct a variety of interactive visualizations with little effort. By using these state-of-art tools, we expect visualization education in political science would be much more diverse. Political scientists can also be assisted in their data analysis procedures from different perspectives through visualization. Furthermore, this will eventually let political scientists recognize the limitations of existing visualization techniques and demand new visualization representations, leading to collaboration with visualization researchers.


\section{Conclusion}\label{sec_conclusion}
In this report, we first reviewed visualization techniques and applications designed for political scientists in the field of data visualization. We found 37 relevant works and categorized them by the political scientists' analysis goals and visualization types. Following that, we reviewed the visualization applications and reported the tasks of domain experts. Based on the tasks, we summarized the system design requirements. Next, we looked at how visualization techniques are used in the political science field, as well as how researchers think about data visualization. They tended to use more visualizations in publications these days, but only a few visualization methods were used for result reporting purposes. Finally, we discussed the knowledge gap in data visualization and visual analytics between the two fields. This work is expected to ease the gap and promote interdisciplinary work between researchers in the field of data visualization and political science.

\ifCLASSOPTIONcaptionsoff
  \newpage
\fi



\bibliographystyle{IEEEtran}

\bibliography{0_main.bib}

\begin{thebibliography}{100}
\providecommand{\url}[1]{#1}
\csname url@samestyle\endcsname
\providecommand{\newblock}{\relax}
\providecommand{\bibinfo}[2]{#2}
\providecommand{\BIBentrySTDinterwordspacing}{\spaceskip=0pt\relax}
\providecommand{\BIBentryALTinterwordstretchfactor}{4}
\providecommand{\BIBentryALTinterwordspacing}{\spaceskip=\fontdimen2\font plus
\BIBentryALTinterwordstretchfactor\fontdimen3\font minus \fontdimen4\font\relax}
\providecommand{\BIBforeignlanguage}[2]{{%
\expandafter\ifx\csname l@#1\endcsname\relax
\typeout{** WARNING: IEEEtran.bst: No hyphenation pattern has been}%
\typeout{** loaded for the language `#1'. Using the pattern for}%
\typeout{** the default language instead.}%
\else
\language=\csname l@#1\endcsname
\fi
#2}}
\providecommand{\BIBdecl}{\relax}
\BIBdecl

\bibitem{dork2013critical}
M.~D{\"o}rk, P.~Feng, C.~Collins, and S.~Carpendale, ``Critical infovis: exploring the politics of visualization,'' in \emph{CHI'13 Extended Abstracts on Human Factors in Computing Systems}, 2013, pp. 2189--2198.

\bibitem{traunmuller2020visualizing}
R.~Traunm{\"u}ller, ``Visualizing data in political science,'' \emph{The SAGE Handbook of Research Methods in Political Science and International Relations}, p. 436, 2020.

\bibitem{segel2010narrative}
E.~Segel and J.~Heer, ``Narrative visualization: Telling stories with data,'' \emph{IEEE transactions on visualization and computer graphics}, vol.~16, no.~6, pp. 1139--1148, 2010.

\bibitem{friendly2001milestones}
M.~Friendly and D.~J. Denis, ``Milestones in the history of thematic cartography, statistical graphics, and data visualization,'' \emph{URL http://www. datavis. ca/milestones}, vol.~32, p.~13, 2001.

\bibitem{election_nyt}
\BIBentryALTinterwordspacing
``An extremely detailed map of the 2020 election,'' accessed: 2022-02-01. [Online]. Available: \url{https://www.nytimes.com/interactive/2021/upshot/2020-election-map.html}
\BIBentrySTDinterwordspacing

\bibitem{cao2012whisper}
N.~Cao, Y.-R. Lin, X.~Sun, D.~Lazer, S.~Liu, and H.~Qu, ``Whisper: Tracing the spatiotemporal process of information diffusion in real time,'' \emph{IEEE transactions on visualization and computer graphics}, vol.~18, no.~12, pp. 2649--2658, 2012.

\bibitem{wang2012si}
X.~Wang, W.~Dou, Z.~Ma, J.~Villalobos, Y.~Chen, T.~Kraft, and W.~Ribarsky, ``I-si: Scalable architecture for analyzing latent topical-level information from social media data,'' in \emph{Computer Graphics Forum}, vol.~31, no. 3pt4.\hskip 1em plus 0.5em minus 0.4em\relax Wiley Online Library, 2012, pp. 1275--1284.

\bibitem{zinovyev2010data}
\BIBentryALTinterwordspacing
A.~Zinovyev, ``Data visualization in political and social sciences,'' Thousand Oaks, pp. 538--545, Oct 2011, data Visualization. [Online]. Available: \url{https://sk.sagepub.com/reference/intlpoliticalscience}
\BIBentrySTDinterwordspacing

\bibitem{kastellec2007using}
J.~P. Kastellec and E.~L. Leoni, ``Using graphs instead of tables in political science,'' \emph{Perspectives on politics}, vol.~5, no.~4, pp. 755--771, 2007.

\bibitem{amar2005low}
R.~Amar, J.~Eagan, and J.~Stasko, ``Low-level components of analytic activity in information visualization,'' in \emph{IEEE Symposium on Information Visualization, 2005. INFOVIS 2005.}\hskip 1em plus 0.5em minus 0.4em\relax IEEE, 2005, pp. 111--117.

\bibitem{playfair1801commercial}
W.~Playfair, \emph{The commercial and political atlas: representing, by means of stained copper-plate charts, the progress of the commerce, revenues, expenditure and debts of england during the whole of the eighteenth century}.\hskip 1em plus 0.5em minus 0.4em\relax T. Burton, 1801.

\bibitem{tufte:2001}
E.~R. Tufte, ``\BIBforeignlanguage{eng}{The visual display of quantitative information},'' Cheshire, Conn, 2001.

\bibitem{batch2017interactive}
A.~Batch and N.~Elmqvist, ``The interactive visualization gap in initial exploratory data analysis,'' \emph{IEEE transactions on visualization and computer graphics}, vol.~24, no.~1, pp. 278--287, 2017.

\bibitem{isenberg2016vispubdata}
P.~Isenberg, F.~Heimerl, S.~Koch, T.~Isenberg, P.~Xu, C.~D. Stolper, M.~Sedlmair, J.~Chen, T.~M{\"o}ller, and J.~Stasko, ``vispubdata. org: A metadata collection about ieee visualization (vis) publications,'' \emph{IEEE transactions on visualization and computer graphics}, vol.~23, no.~9, pp. 2199--2206, 2016.

\bibitem{hsu2017community}
Y.-C. Hsu, P.~Dille, J.~Cross, B.~Dias, R.~Sargent, and I.~Nourbakhsh, ``Community-empowered air quality monitoring system,'' in \emph{Proceedings of the 2017 CHI Conference on human factors in computing systems}, 2017, pp. 1607--1619.

\bibitem{wan2014improving}
S.~Wan and C.~Paris, ``Improving government services with social media feedback,'' in \emph{Proceedings of the 19th international conference on Intelligent User Interfaces}, 2014, pp. 27--36.

\bibitem{tariq2021planning}
Z.~Tariq, M.~Mannino, M.~Le~Xuan~Anh, W.~Bagge, A.~Abouzied, and D.~Shasha, ``Planning epidemic interventions with epipolicy,'' in \emph{The 34th Annual ACM Symposium on User Interface Software and Technology}, 2021, pp. 894--909.

\bibitem{livnat2012epinome}
Y.~Livnat, T.-M. Rhyne, and M.~Samore, ``Epinome: A visual-analytics workbench for epidemiology data,'' \emph{IEEE computer graphics and applications}, vol.~32, no.~2, pp. 89--95, 2012.

\bibitem{chen2017social}
S.~Chen, L.~Lin, and X.~Yuan, ``Social media visual analytics,'' in \emph{Computer Graphics Forum}, vol.~36, no.~3.\hskip 1em plus 0.5em minus 0.4em\relax Wiley Online Library, 2017, pp. 563--587.

\bibitem{wu2016survey}
Y.~Wu, N.~Cao, D.~Gotz, Y.-P. Tan, and D.~A. Keim, ``A survey on visual analytics of social media data,'' \emph{IEEE Transactions on Multimedia}, vol.~18, no.~11, pp. 2135--2148, 2016.

\bibitem{cashman2020cava}
D.~Cashman, S.~Xu, S.~Das, F.~Heimerl, C.~Liu, S.~R. Humayoun, M.~Gleicher, A.~Endert, and R.~Chang, ``Cava: A visual analytics system for exploratory columnar data augmentation using knowledge graphs,'' \emph{IEEE Transactions on Visualization and Computer Graphics}, vol.~27, no.~2, pp. 1731--1741, 2020.

\bibitem{shneiderman2006network}
B.~Shneiderman and A.~Aris, ``Network visualization by semantic substrates,'' \emph{IEEE transactions on visualization and computer graphics}, vol.~12, no.~5, pp. 733--740, 2006.

\bibitem{perer2006balancing}
A.~Perer and B.~Shneiderman, ``Balancing systematic and flexible exploration of social networks,'' \emph{IEEE transactions on visualization and computer graphics}, vol.~12, no.~5, pp. 693--700, 2006.

\bibitem{dunne2013motif}
C.~Dunne and B.~Shneiderman, ``Motif simplification: improving network visualization readability with fan, connector, and clique glyphs,'' in \emph{Proceedings of the SIGCHI Conference on Human Factors in Computing Systems}, 2013, pp. 3247--3256.

\bibitem{chan2018v}
G.~Y.-Y. Chan, P.~Xu, Z.~Dai, and L.~Ren, ``V i b r: Visualizing bipartite relations at scale with the minimum description length principle,'' \emph{IEEE transactions on visualization and computer graphics}, vol.~25, no.~1, pp. 321--330, 2018.

\bibitem{SPID_dataset}
\BIBentryALTinterwordspacing
``Spid: State policy innovation and diffusion database dataverse,'' accessed: 2022-02-01. [Online]. Available: \url{https://dataverse.harvard.edu/dataverse/spid}
\BIBentrySTDinterwordspacing

\bibitem{FAOSTAT}
\BIBentryALTinterwordspacing
``Food and agriculture organization of united nations,'' accessed: 2022-02-01. [Online]. Available: \url{https://www.fao.org/faostat/en/}
\BIBentrySTDinterwordspacing

\bibitem{WGI}
\BIBentryALTinterwordspacing
``Food and agriculture organization of united nations,'' accessed: 2022-02-01. [Online]. Available: \url{http://info.worldbank.org/governance/wgi/Home/Reports}
\BIBentrySTDinterwordspacing

\bibitem{theWorldBank}
\BIBentryALTinterwordspacing
``The world bank,'' accessed: 2022-02-01. [Online]. Available: \url{https://databank.worldbank.org/home.aspx}
\BIBentrySTDinterwordspacing

\bibitem{raleigh2010introducing}
C.~Raleigh, A.~Linke, H.~Hegre, and J.~Karlsen, ``Introducing acled: an armed conflict location and event dataset: special data feature,'' \emph{Journal of peace research}, vol.~47, no.~5, pp. 651--660, 2010.

\bibitem{HarvardDataverse}
\BIBentryALTinterwordspacing
``Harvard dataverse,'' accessed: 2022-02-01. [Online]. Available: \url{https://dataverse.harvard.edu/}
\BIBentrySTDinterwordspacing

\bibitem{michigan_finding_data}
\BIBentryALTinterwordspacing
``University of michigan library finding data,'' accessed: 2022-02-01. [Online]. Available: \url{https://guides.lib.umich.edu/findingdata}
\BIBentrySTDinterwordspacing

\bibitem{michigan_finding_statistics}
\BIBentryALTinterwordspacing
``University of michigan library finding statistics and data sets,'' accessed: 2022-02-01. [Online]. Available: \url{https://guides.lib.umich.edu/govstatistics}
\BIBentrySTDinterwordspacing

\bibitem{data_gov}
\BIBentryALTinterwordspacing
``U.s. government’s open data,'' accessed: 2022-02-01. [Online]. Available: \url{https://catalog.data.gov/dataset}
\BIBentrySTDinterwordspacing

\bibitem{brandes2006summarizing}
U.~Brandes, D.~Fleischer, and J.~Lerner, ``Summarizing dynamic bipolar conflict structures,'' \emph{IEEE Transactions on Visualization and Computer Graphics}, vol.~12, no.~6, pp. 1486--1499, 2006.

\bibitem{shen2006visual}
Z.~Shen, K.-L. Ma, and T.~Eliassi-Rad, ``Visual analysis of large heterogeneous social networks by semantic and structural abstraction,'' \emph{IEEE transactions on visualization and computer graphics}, vol.~12, no.~6, pp. 1427--1439, 2006.

\bibitem{wang2008investigative}
X.~Wang, E.~Miller, K.~Smarick, W.~Ribarsky, and R.~Chang, ``Investigative visual analysis of global terrorism,'' in \emph{Computer Graphics Forum}, vol.~27, no.~3.\hskip 1em plus 0.5em minus 0.4em\relax Wiley Online Library, 2008, pp. 919--926.

\bibitem{reda2011visualizing}
K.~Reda, C.~Tantipathananandh, A.~Johnson, J.~Leigh, and T.~Berger-Wolf, ``Visualizing the evolution of community structures in dynamic social networks,'' in \emph{Computer Graphics Forum}, vol.~30, no.~3.\hskip 1em plus 0.5em minus 0.4em\relax Wiley Online Library, 2011, pp. 1061--1070.

\bibitem{dang2016timearcs}
T.~N. Dang, N.~Pendar, and A.~G. Forbes, ``Timearcs: Visualizing fluctuations in dynamic networks,'' in \emph{Computer Graphics Forum}, vol.~35, no.~3.\hskip 1em plus 0.5em minus 0.4em\relax Wiley Online Library, 2016, pp. 61--69.

\bibitem{wang2018visual}
H.~Wang, Y.~Lu, S.~T. Shutters, M.~Steptoe, F.~Wang, S.~Landis, and R.~Maciejewski, ``A visual analytics framework for spatiotemporal trade network analysis,'' \emph{IEEE transactions on visualization and computer graphics}, vol.~25, no.~1, pp. 331--341, 2018.

\bibitem{dinkla2015dual}
K.~Dinkla, N.~H. Riche, and M.~A. Westenberg, ``Dual adjacency matrix: exploring link groups in dense networks,'' in \emph{Computer Graphics Forum}, vol.~34, no.~3.\hskip 1em plus 0.5em minus 0.4em\relax Wiley Online Library, 2015, pp. 311--320.

\bibitem{lu2017visual}
Y.~Lu, H.~Wang, S.~Landis, and R.~Maciejewski, ``A visual analytics framework for identifying topic drivers in media events,'' \emph{IEEE transactions on visualization and computer graphics}, vol.~24, no.~9, pp. 2501--2515, 2017.

\bibitem{el2017nerex}
M.~El-Assady, R.~Sevastjanova, B.~Gipp, D.~Keim, and C.~Collins, ``Nerex: Named-entity relationship exploration in multi-party conversations,'' in \emph{Computer Graphics Forum}, vol.~36, no.~3.\hskip 1em plus 0.5em minus 0.4em\relax Wiley Online Library, 2017, pp. 213--225.

\bibitem{el2018visual}
M.~El-Assady, F.~Sperrle, O.~Deussen, D.~Keim, and C.~Collins, ``Visual analytics for topic model optimization based on user-steerable speculative execution,'' \emph{IEEE transactions on visualization and computer graphics}, vol.~25, no.~1, pp. 374--384, 2018.

\bibitem{el2018threadreconstructor}
M.~El-Assady, R.~Sevastjanova, D.~Keim, and C.~Collins, ``Threadreconstructor: Modeling reply-chains to untangle conversational text through visual analytics,'' in \emph{Computer Graphics Forum}, vol.~37, no.~3.\hskip 1em plus 0.5em minus 0.4em\relax Wiley Online Library, 2018, pp. 351--365.

\bibitem{cui2014hierarchical}
W.~Cui, S.~Liu, Z.~Wu, and H.~Wei, ``How hierarchical topics evolve in large text corpora,'' \emph{IEEE transactions on visualization and computer graphics}, vol.~20, no.~12, pp. 2281--2290, 2014.

\bibitem{el2016contovi}
M.~El-Assady, V.~Gold, C.~Acevedo, C.~Collins, and D.~Keim, ``Contovi: Multi-party conversation exploration using topic-space views,'' in \emph{Computer Graphics Forum}, vol.~35, no.~3.\hskip 1em plus 0.5em minus 0.4em\relax Wiley Online Library, 2016, pp. 431--440.

\bibitem{ahn2020policyflow}
Y.~Ahn and Y.-R. Lin, ``Policyflow: Interpreting policy diffusion in context,'' \emph{ACM Transactions on Interactive Intelligent Systems (TiiS)}, vol.~10, no.~2, pp. 1--23, 2020.

\bibitem{habermas1985theory}
J.~Habermas, \emph{The theory of communicative action: Volume 1: Reason and the rationalization of society}.\hskip 1em plus 0.5em minus 0.4em\relax Beacon press, 1985, vol.~1.

\bibitem{persson2013effects}
M.~Persson, P.~Esaiasson, and M.~Gilljam, ``The effects of direct voting and deliberation on legitimacy beliefs: An experimental study of small group decision-making,'' \emph{European Political Science Review}, vol.~5, no.~3, pp. 381--399, 2013.

\bibitem{dorling2006worldmapper}
D.~Dorling, A.~Barford, and M.~Newman, ``Worldmapper: The world as you've never seen it before,'' \emph{IEEE transactions on visualization and computer graphics}, vol.~12, no.~5, pp. 757--764, 2006.

\bibitem{nusrat2017cartogram}
S.~Nusrat, M.~J. Alam, C.~Scheidegger, and S.~Kobourov, ``Cartogram visualization for bivariate geo-statistical data,'' \emph{IEEE transactions on visualization and computer graphics}, vol.~24, no.~10, pp. 2675--2688, 2017.

\bibitem{li2018concavecubes}
M.~Li, F.~Choudhury, Z.~Bao, H.~Samet, and T.~Sellis, ``Concavecubes: Supporting cluster-based geographical visualization in large data scale,'' in \emph{Computer Graphics Forum}, vol.~37, no.~3.\hskip 1em plus 0.5em minus 0.4em\relax Wiley Online Library, 2018, pp. 217--228.

\bibitem{turkay2014attribute}
C.~Turkay, A.~Slingsby, H.~Hauser, J.~Wood, and J.~Dykes, ``Attribute signatures: Dynamic visual summaries for analyzing multivariate geographical data,'' \emph{IEEE Transactions on Visualization and Computer Graphics}, vol.~20, no.~12, pp. 2033--2042, 2014.

\bibitem{yates2014visualizing}
A.~Yates, A.~Webb, M.~Sharpnack, H.~Chamberlin, K.~Huang, and R.~Machiraju, ``Visualizing multidimensional data with glyph sploms,'' in \emph{Computer Graphics Forum}, vol.~33, no.~3.\hskip 1em plus 0.5em minus 0.4em\relax Wiley Online Library, 2014, pp. 301--310.

\bibitem{goodwin2015visualizing}
S.~Goodwin, J.~Dykes, A.~Slingsby, and C.~Turkay, ``Visualizing multiple variables across scale and geography,'' \emph{IEEE Transactions on Visualization and Computer Graphics}, vol.~22, no.~1, pp. 599--608, 2015.

\bibitem{cui2011textflow}
W.~Cui, S.~Liu, L.~Tan, C.~Shi, Y.~Song, Z.~Gao, H.~Qu, and X.~Tong, ``Textflow: Towards better understanding of evolving topics in text,'' \emph{IEEE transactions on visualization and computer graphics}, vol.~17, no.~12, pp. 2412--2421, 2011.

\bibitem{sun2014evoriver}
G.~Sun, Y.~Wu, S.~Liu, T.-Q. Peng, J.~J. Zhu, and R.~Liang, ``Evoriver: Visual analysis of topic coopetition on social media,'' \emph{IEEE transactions on visualization and computer graphics}, vol.~20, no.~12, pp. 1753--1762, 2014.

\bibitem{slingsby2011exploring}
A.~Slingsby, J.~Dykes, and J.~Wood, ``Exploring uncertainty in geodemographics with interactive graphics,'' \emph{IEEE Transactions on Visualization and Computer Graphics}, vol.~17, no.~12, pp. 2545--2554, 2011.

\bibitem{huang2019exploring}
Z.~Huang, Y.~Lu, E.~A. Mack, W.~Chen, and R.~Maciejewski, ``Exploring the sensitivity of choropleths under attribute uncertainty,'' \emph{IEEE Transactions on Visualization and Computer Graphics}, vol.~26, no.~8, pp. 2576--2590, 2019.

\bibitem{draper2008votes}
G.~Draper and R.~Riesenfeld, ``Who votes for what? a visual query language for opinion data,'' \emph{IEEE transactions on visualization and computer graphics}, vol.~14, no.~6, pp. 1197--1204, 2008.

\bibitem{heilmann2004recmap}
R.~Heilmann, D.~A. Keim, C.~Panse, and M.~Sips, ``Recmap: Rectangular map approximations,'' in \emph{IEEE Symposium on Information Visualization}.\hskip 1em plus 0.5em minus 0.4em\relax IEEE, 2004, pp. 33--40.

\bibitem{cano2015mosaic}
R.~G. Cano, K.~Buchin, T.~Castermans, A.~Pieterse, W.~Sonke, and B.~Speckmann, ``Mosaic drawings and cartograms,'' in \emph{Computer Graphics Forum}, vol.~34, no.~3.\hskip 1em plus 0.5em minus 0.4em\relax Wiley Online Library, 2015, pp. 361--370.

\bibitem{hografer2020state}
M.~Hogr{\"a}fer, M.~Heitzler, and H.-J. Schulz, ``The state of the art in map-like visualization,'' in \emph{Computer Graphics Forum}, vol.~39, no.~3.\hskip 1em plus 0.5em minus 0.4em\relax Wiley Online Library, 2020, pp. 647--674.

\bibitem{guo2009flow}
D.~Guo, ``Flow mapping and multivariate visualization of large spatial interaction data,'' \emph{IEEE Transactions on Visualization and Computer Graphics}, vol.~15, no.~6, pp. 1041--1048, 2009.

\bibitem{speckmann2010necklace}
B.~Speckmann and K.~Verbeek, ``Necklace maps.'' \emph{IEEE Trans. Vis. Comput. Graph.}, vol.~16, no.~6, pp. 881--889, 2010.

\bibitem{guo2014origin}
D.~Guo and X.~Zhu, ``Origin-destination flow data smoothing and mapping,'' \emph{IEEE transactions on visualization and computer graphics}, vol.~20, no.~12, pp. 2043--2052, 2014.

\bibitem{boyandin2011flowstrates}
I.~Boyandin, E.~Bertini, P.~Bak, and D.~Lalanne, ``Flowstrates: An approach for visual exploration of temporal origin-destination data,'' in \emph{Computer Graphics Forum}, vol.~30, no.~3.\hskip 1em plus 0.5em minus 0.4em\relax Wiley Online Library, 2011, pp. 971--980.

\bibitem{yang2016many}
Y.~Yang, T.~Dwyer, S.~Goodwin, and K.~Marriott, ``Many-to-many geographically-embedded flow visualisation: An evaluation,'' \emph{IEEE transactions on visualization and computer graphics}, vol.~23, no.~1, pp. 411--420, 2016.

\bibitem{van2014multivariate}
S.~Van~den Elzen and J.~J. Van~Wijk, ``Multivariate network exploration and presentation: From detail to overview via selections and aggregations,'' \emph{IEEE Transactions on Visualization and Computer Graphics}, vol.~20, no.~12, pp. 2310--2319, 2014.

\bibitem{ChicagoDataPortal}
\BIBentryALTinterwordspacing
``Chicago data portal,'' accessed: 2023-04-05. [Online]. Available: \url{https://data.cityofchicago.org/}
\BIBentrySTDinterwordspacing

\bibitem{cox19963d}
K.~C. Cox, S.~G. Eick, and T.~He, ``3d geographic network displays,'' \emph{ACM Sigmod Record}, vol.~25, no.~4, pp. 50--54, 1996.

\bibitem{munzner1996visualizing}
T.~Munzner, E.~Hoffman, K.~Claffy, and B.~Fenner, ``Visualizing the global topology of the mbone,'' in \emph{Proceedings IEEE Symposium on Information Visualization'96}.\hskip 1em plus 0.5em minus 0.4em\relax IEEE, 1996, pp. 85--92.

\bibitem{tobler1987experiments}
W.~R. Tobler, ``Experiments in migration mapping by computer,'' \emph{The American Cartographer}, vol.~14, no.~2, pp. 155--163, 1987.

\bibitem{phan2005flow}
D.~Phan, L.~Xiao, R.~Yeh, and P.~Hanrahan, ``Flow map layout,'' in \emph{IEEE Symposium on Information Visualization, 2005. INFOVIS 2005.}\hskip 1em plus 0.5em minus 0.4em\relax IEEE, 2005, pp. 219--224.

\bibitem{buchin2011flow}
K.~Buchin, B.~Speckmann, and K.~Verbeek, ``Flow map layout via spiral trees,'' \emph{IEEE transactions on visualization and computer graphics}, vol.~17, no.~12, pp. 2536--2544, 2011.

\bibitem{jenny2018design}
B.~Jenny, D.~M. Stephen, I.~Muehlenhaus, B.~E. Marston, R.~Sharma, E.~Zhang, and H.~Jenny, ``Design principles for origin-destination flow maps,'' \emph{Cartography and Geographic Information Science}, vol.~45, no.~1, pp. 62--75, 2018.

\bibitem{wood2010visualisation}
J.~Wood, J.~Dykes, and A.~Slingsby, ``Visualisation of origins, destinations and flows with od maps,'' \emph{The Cartographic Journal}, vol.~47, no.~2, pp. 117--129, 2010.

\bibitem{pupyrev2012edge}
S.~Pupyrev, L.~Nachmanson, S.~Bereg, and A.~E. Holroyd, ``Edge routing with ordered bundles,'' in \emph{Graph Drawing: 19th International Symposium, GD 2011, Eindhoven, The Netherlands, September 21-23, 2011, Revised Selected Papers 19}.\hskip 1em plus 0.5em minus 0.4em\relax Springer, 2012, pp. 136--147.

\bibitem{IRS}
\BIBentryALTinterwordspacing
``Internal revenue service, soi tax stats - county-to-county migration data files,'' accessed: 2023-04-05. [Online]. Available: \url{https://www.irs.gov/statistics/soi-tax-stats-county-to-county-migration-data-files}
\BIBentrySTDinterwordspacing

\bibitem{sacha2014knowledge}
D.~Sacha, A.~Stoffel, F.~Stoffel, B.~C. Kwon, G.~Ellis, and D.~A. Keim, ``Knowledge generation model for visual analytics,'' \emph{IEEE transactions on visualization and computer graphics}, vol.~20, no.~12, pp. 1604--1613, 2014.

\bibitem{han2023}
D.~Han, A.-A.-R. Nayeem, J.~Windett, and I.~Cho, ``Pdviz: A visual analytics approach for state policy data,'' in \emph{Computer Graphics Forum}.\hskip 1em plus 0.5em minus 0.4em\relax Wiley Online Library, 2023.

\bibitem{pezzotti2018multiscale}
N.~Pezzotti, J.-D. Fekete, T.~H{\"o}llt, B.~P. Lelieveldt, E.~Eisemann, and A.~Vilanova, ``Multiscale visualization and exploration of large bipartite graphs,'' in \emph{Computer Graphics Forum}, vol.~37, no.~3.\hskip 1em plus 0.5em minus 0.4em\relax Wiley Online Library, 2018, pp. 549--560.

\bibitem{hsiang2011civil}
S.~M. Hsiang, K.~C. Meng, and M.~A. Cane, ``Civil conflicts are associated with the global climate,'' \emph{Nature}, vol. 476, no. 7361, pp. 438--441, 2011.

\bibitem{marshall2008fragility}
M.~G. Marshall, ``Fragility, instability, and the failure of states,'' \emph{Center for}, 2008.

\bibitem{crouser2011two}
R.~J. Crouser, D.~Kee, D.~Jeong, and R.~Chang, ``Two visualization tools for analyzing agent-based simulations in political science,'' \emph{IEEE Computer Graphics and Applications}, vol.~32, no.~1, pp. 67--77, 2011.

\bibitem{lee2016vlat}
S.~Lee, S.-H. Kim, and B.~C. Kwon, ``Vlat: Development of a visualization literacy assessment test,'' \emph{IEEE transactions on visualization and computer graphics}, vol.~23, no.~1, pp. 551--560, 2016.

\bibitem{jones2005beyond}
B.~S. Jones and R.~P. Branton, ``Beyond logit and probit: Cox duration models of single, repeating, and competing events for state policy adoption,'' \emph{State Politics \& Policy Quarterly}, vol.~5, no.~4, pp. 420--443, 2005.

\bibitem{tyranny}
\BIBentryALTinterwordspacing
S.~Saideman, ``The tyranny of the big 3? which journals count most may be increasingly problematic,'' 2018. [Online]. Available: \url{https://www.duckofminerva.com/2018/04/the-tyranny-of-the-big-3-which-journals-count-most-may-be-increasingly-problematic.html}
\BIBentrySTDinterwordspacing

\bibitem{chen2009toward}
Y.~Chen, J.~Yang, and W.~Ribarsky, ``Toward effective insight management in visual analytics systems,'' in \emph{2009 IEEE Pacific Visualization Symposium}.\hskip 1em plus 0.5em minus 0.4em\relax IEEE, 2009, pp. 49--56.

\bibitem{AJPS}
\BIBentryALTinterwordspacing
``The american journal of political science,'' accessed: 2022-02-01. [Online]. Available: \url{"https://ajps.org/",}
\BIBentrySTDinterwordspacing

\bibitem{visbest}
\BIBentryALTinterwordspacing
``The jpr best visualization awards,'' accessed: 2022-02-01. [Online]. Available: \url{https://journals.sagepub.com/page/jpr/visualization-awards/index}
\BIBentrySTDinterwordspacing

\bibitem{ito2021does}
G.~Ito, ``Why does ethnic partition foster violence? unpacking the deep historical roots of civil conflicts,'' \emph{Journal of Peace Research}, vol.~58, no.~5, pp. 986--1003, 2021.

\bibitem{anders2020territorial}
T.~Anders, ``Territorial control in civil wars: Theory and measurement using machine learning,'' \emph{Journal of Peace Research}, vol.~57, no.~6, pp. 701--714, 2020.

\bibitem{van2019climate}
S.~van Weezel, ``On climate and conflict: Precipitation decline and communal conflict in ethiopia and kenya,'' \emph{Journal of Peace Research}, vol.~56, no.~4, pp. 514--528, 2019.

\bibitem{muller2018new}
C.~M{\"u}ller-Crepon and P.~Hunziker, ``New spatial data on ethnicity: Introducing side,'' \emph{Journal of Peace Research}, vol.~55, no.~5, pp. 687--698, 2018.

\bibitem{colaresi2017robot}
M.~Colaresi and Z.~Mahmood, ``Do the robot: Lessons from machine learning to improve conflict forecasting,'' \emph{Journal of Peace Research}, vol.~54, no.~2, pp. 193--214, 2017.

\bibitem{kinne2016agreeing}
B.~J. Kinne, ``Agreeing to arm: Bilateral weapons agreements and the global arms trade,'' \emph{Journal of Peace Research}, vol.~53, no.~3, pp. 359--377, 2016.

\bibitem{baum2015filtering}
M.~A. Baum and Y.~M. Zhukov, ``Filtering revolution: Reporting bias in international newspaper coverage of the libyan civil war,'' \emph{Journal of Peace Research}, vol.~52, no.~3, pp. 384--400, 2015.

\bibitem{chadefaux2014early}
T.~Chadefaux, ``Early warning signals for war in the news,'' \emph{Journal of Peace Research}, vol.~51, no.~1, pp. 5--18, 2014.

\bibitem{kinne2013igo}
B.~J. Kinne, ``Igo membership, network convergence, and credible signaling in militarized disputes,'' \emph{Journal of Peace Research}, vol.~50, no.~6, pp. 659--676, 2013.

\bibitem{USNEWSPOL}
\BIBentryALTinterwordspacing
``the us news and reports political science rankings,'' accessed: 2022-02-01. [Online]. Available: \url{"https://www.usnews.com/best-graduate-schools/top-humanities-schools/american-politics-rankings",}
\BIBentrySTDinterwordspacing

\bibitem{oprisko2012}
R.~Oprisko, ``Superpowers: The american academic elite,'' \emph{Georgetown Public Policy Review}, 2012.

\bibitem{USNEWSMET}
\BIBentryALTinterwordspacing
``the us news and reports political methodology rankings,'' accessed: 2022-02-01. [Online]. Available: \url{"https://www.usnews.com/best-graduate-schools/top-humanities-schools/political-methodology-rankings",}
\BIBentrySTDinterwordspacing

\bibitem{borkin2013makes}
M.~A. Borkin, A.~A. Vo, Z.~Bylinskii, P.~Isola, S.~Sunkavalli, A.~Oliva, and H.~Pfister, ``What makes a visualization memorable?'' \emph{IEEE transactions on visualization and computer graphics}, vol.~19, no.~12, pp. 2306--2315, 2013.

\bibitem{matejka2017same}
J.~Matejka and G.~Fitzmaurice, ``Same stats, different graphs: generating datasets with varied appearance and identical statistics through simulated annealing,'' in \emph{Proceedings of the 2017 CHI conference on human factors in computing systems}, 2017, pp. 1290--1294.

\bibitem{xu2018chart}
S.~Xu, C.~Bryan, J.~K. Li, J.~Zhao, and K.-L. Ma, ``Chart constellations: Effective chart summarization for collaborative and multi-user analyses,'' in \emph{Computer Graphics Forum}, vol.~37, no.~3.\hskip 1em plus 0.5em minus 0.4em\relax Wiley Online Library, 2018, pp. 75--86.

\bibitem{li2020resolving}
J.~K. Li, S.~Xu, Y.~Ye, and K.-L. Ma, ``Resolving conflicting insights in asynchronous collaborative visual analysis,'' in \emph{Computer Graphics Forum}, vol.~39, no.~3.\hskip 1em plus 0.5em minus 0.4em\relax Wiley Online Library, 2020, pp. 497--509.

\bibitem{ArcGISBlog}
\BIBentryALTinterwordspacing
``Arcgis blog,'' accessed: 2023-04-05. [Online]. Available: \url{https://www.esri.com/arcgis-blog/overview/}
\BIBentrySTDinterwordspacing

\bibitem{harrie2021thematic}
L.~Harrie, ``Thematic mapping: 101 inspiring ways to visualise empirical data: by kenneth fieldesri press, redlands, 2022, 296 pp., usd 59.99 (pbk), isbn 9781589485570,'' 2021.

\bibitem{ComparativeAgendasProject}
\BIBentryALTinterwordspacing
``Comparative agendas project,'' accessed: 2023-04-05. [Online]. Available: \url{https://www.comparativeagendas.net/}
\BIBentrySTDinterwordspacing

\bibitem{VoteView}
\BIBentryALTinterwordspacing
``Vote view blog,'' accessed: 2023-04-05. [Online]. Available: \url{https://voteview.com/rollcall/RS1180081}
\BIBentrySTDinterwordspacing

\bibitem{CSPP}
\BIBentryALTinterwordspacing
``Cspp,'' accessed: 2023-04-05. [Online]. Available: \url{https://cspp.ippsr.msu.edu/cspp/}
\BIBentrySTDinterwordspacing

\bibitem{raleigh:etal:2010}
\BIBentryALTinterwordspacing
C.~Raleigh, A.~Linke, H.~Hegre, and J.~Karlsen, ``Introducing acled-armed conflict location and event data,'' \emph{Journal of Peace Research}, vol.~47, pp. 651--660, 2010. [Online]. Available: \url{https://acleddata.com/dashboard/dashboard}
\BIBentrySTDinterwordspacing

\bibitem{tollefsen:etal:2012}
\BIBentryALTinterwordspacing
A.~F. Tollefsen, H.~Strand, and H.~Buhaug, ``Prio-grid: A unified spatial data structure,'' \emph{Journal of Peace Research}, vol.~49, pp. 363--374, 2012. [Online]. Available: \url{https://grid.prio.org/}
\BIBentrySTDinterwordspacing

\bibitem{merz:etal:2016}
\BIBentryALTinterwordspacing
N.~Merz, S.~Regel, and J.~Lewandowski, ``The manifesto corpus: A new resource for research on political parties and quantitative text analysis,'' \emph{Research \& Politics}, vol.~3, no.~2, p. 2053168016643346, 2016. [Online]. Available: \url{https://manifesto-project.wzb.eu/information/documents/visualizations}
\BIBentrySTDinterwordspacing

\bibitem{dasgupta2016familiarity}
A.~Dasgupta, J.-Y. Lee, R.~Wilson, R.~A. Lafrance, N.~Cramer, K.~Cook, and S.~Payne, ``Familiarity vs trust: A comparative study of domain scientists' trust in visual analytics and conventional analysis methods,'' \emph{IEEE transactions on visualization and computer graphics}, vol.~23, no.~1, pp. 271--280, 2016.

\bibitem{boukhelifa2009uncertainty}
N.~Boukhelifa and D.~J. Duke, ``Uncertainty visualization: why might it fail?'' in \emph{CHI'09 extended abstracts on human factors in computing systems}, 2009, pp. 4051--4056.

\bibitem{greis2017designing}
M.~Greis, J.~Hullman, M.~Correll, M.~Kay, and O.~Shaer, ``Designing for uncertainty in hci: When does uncertainty help?'' in \emph{Proceedings of the 2017 CHI conference extended abstracts on human factors in computing systems}, 2017, pp. 593--600.

\bibitem{wesslen2021effect}
R.~Wesslen, A.~Karduni, D.~Markant, and W.~Dou, ``Effect of uncertainty visualizations on myopic loss aversion and the equity premium puzzle in retirement investment decisions,'' \emph{IEEE Transactions on Visualization and Computer Graphics}, vol.~28, no.~1, pp. 454--464, 2021.

\bibitem{sacha2015role}
D.~Sacha, H.~Senaratne, B.~C. Kwon, G.~Ellis, and D.~A. Keim, ``The role of uncertainty, awareness, and trust in visual analytics,'' \emph{IEEE transactions on visualization and computer graphics}, vol.~22, no.~1, pp. 240--249, 2015.

\bibitem{ggplot}
\BIBentryALTinterwordspacing
H.~Wickham, \emph{ggplot2: Elegant Graphics for Data Analysis}.\hskip 1em plus 0.5em minus 0.4em\relax Springer-Verlag New York, 2016. [Online]. Available: \url{https://ggplot2.tidyverse.org}
\BIBentrySTDinterwordspacing

\bibitem{rShiney}
\BIBentryALTinterwordspacing
{R Shiney}, accessed: 2023-2-9. [Online]. Available: \url{https://shiny.rstudio.com/}
\BIBentrySTDinterwordspacing

\bibitem{Stata}
\BIBentryALTinterwordspacing
{STATA}, accessed: 2023-2-9. [Online]. Available: \url{https://www.stata.com/}
\BIBentrySTDinterwordspacing

\bibitem{Excel}
\BIBentryALTinterwordspacing
{Excel}, accessed: 2023-2-9. [Online]. Available: \url{https://www.microsoft.com/en-us/microsoft-365/excel}
\BIBentrySTDinterwordspacing

\bibitem{Tableau}
\BIBentryALTinterwordspacing
{Tableau}, accessed: 2023-2-9. [Online]. Available: \url{https://dataverse.harvard.edu/dataverse/spid}
\BIBentrySTDinterwordspacing

\bibitem{Spotfire}
\BIBentryALTinterwordspacing
{TIBCO Spotfire}, accessed: 2023-2-9. [Online]. Available: \url{https://www.tibco.com/products/tibco-spotfire}
\BIBentrySTDinterwordspacing

\bibitem{ADVIZOR}
\BIBentryALTinterwordspacing
{ADVIZOR}, accessed: 2023-2-9. [Online]. Available: \url{https://www.advizorsolutions.com/}
\BIBentrySTDinterwordspacing

\bibitem{boyandin2010using}
I.~Boyandin, E.~Bertini, and D.~Lalanne, ``Using flow maps to explore migrations over time,'' in \emph{Geospatial Visual Analytics Workshop in conjunction with The 13th AGILE International Conference on Geographic Information Science}, vol.~2, no.~3, 2010.

\bibitem{alles2021burden}
S.~Alles, M.~Pach{\'o}n, and M.~Mu{\~n}oz, ``The burden of election logistics: Election ballots and the territorial influence of party machines in colombia,'' \emph{The Journal of Politics}, vol.~83, no.~4, pp. 1635--1651, 2021.

\bibitem{dinesen2021legislators}
P.~T. Dinesen, M.~Dahl, and M.~Schi{\o}ler, ``When are legislators responsive to ethnic minorities? testing the role of electoral incentives and candidate selection for mitigating ethnocentric responsiveness,'' \emph{American Political Science Review}, vol. 115, no.~2, pp. 450--466, 2021.

\bibitem{boggild2021citizens}
T.~B{\o}ggild, L.~Aar{\o}e, and M.~B. Petersen, ``Citizens as complicits: distrust in politicians and biased social dissemination of political information,'' \emph{American Political Science Review}, vol. 115, no.~1, pp. 269--285, 2021.

\bibitem{eck2021evade}
K.~Eck, S.~Hatz, C.~Crabtree, and A.~Tago, ``Evade and deceive? citizen responses to surveillance,'' \emph{The Journal of Politics}, vol.~83, no.~4, pp. 1545--1558, 2021.

\end{thebibliography}
\nocite{anders2020territorial}
\nocite{van2019climate}
\nocite{muller2018new}
\nocite{colaresi2017robot}
\nocite{kinne2016agreeing}
\nocite{baum2015filtering}
\nocite{chadefaux2014early}
\nocite{kinne2013igo}

\nocite{speckmann2010necklace}
\nocite{munzner1996visualizing}
\nocite{guo2009flow}
\nocite{boyandin2010using}
\nocite{tobler1987experiments}


\nocite{wood2010visualisation}
\nocite{yang2016many}
\nocite{boyandin2011flowstrates}

\nocite{alles2021burden}
\nocite{dinesen2021legislators}
\nocite{boggild2021citizens}
\nocite{eck2021evade}



 \begin{IEEEbiography}
[{\includegraphics[width=1in,height=1.25in,clip,keepaspectratio]{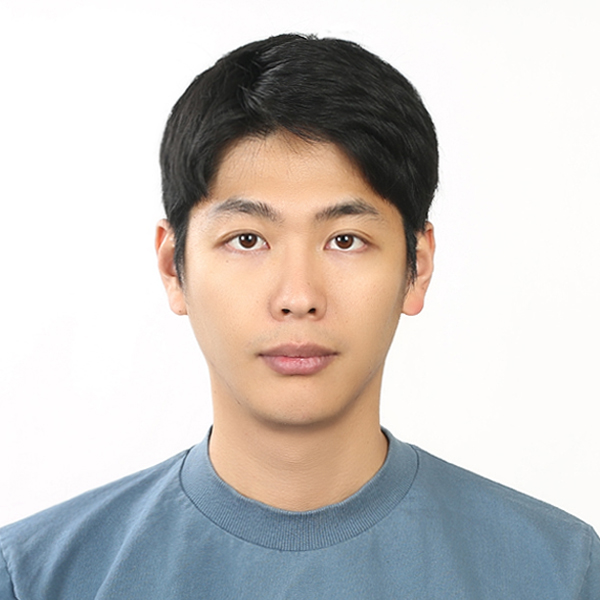}}]{Dongyun Han}
Dongyun Han is a Ph.D. student in the department of Computer Science at Utah State University. His research interests include visual analytics and virtual reality. He has been developing visual analytics systems for several research domains including Earth Science, Political Science, History, and Distributed Energy resources. 
He received an M.S. in Computer Science at UNIST in 2020.   

 \end{IEEEbiography}

 \begin{IEEEbiography}[{\includegraphics[width=1in,height=1.25in,clip,keepaspectratio]{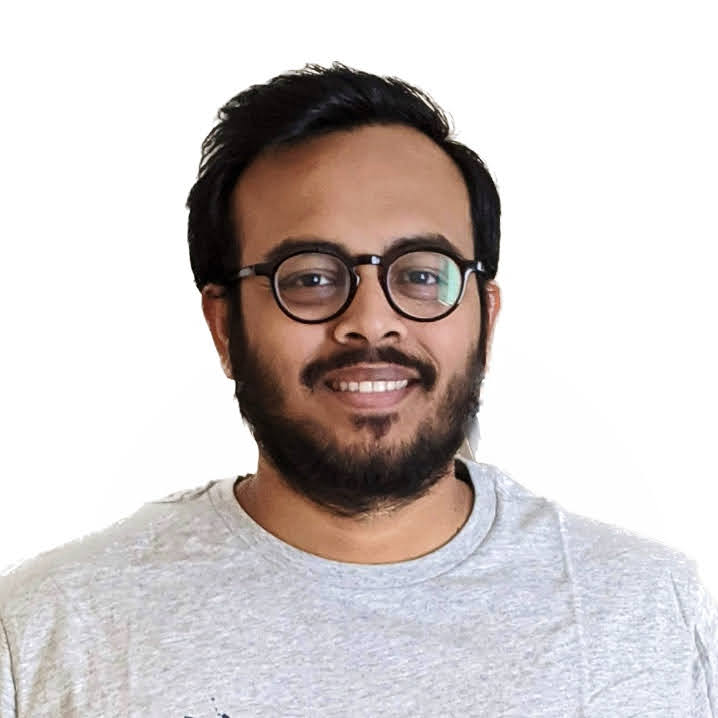}}]{Abdullah-Al-Raihan Nayeem} 
Abdullah-Al-Raihan Nayeem is a research assistant at the Ribarsky Center for Visual Analytics. He received his Ph.D. in Computer Science from the University of North Carolina at Charlotte. 
His research interests are visual analytics systems, interactive geospatial visualizations, and distributed spatiotemporal analytics. 
 \end{IEEEbiography}
 
 \begin{IEEEbiography} [{\includegraphics[width=1in,height=1.25in,clip,keepaspectratio]{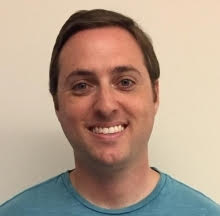}}]{Jason Windett}
Jason Windett is an Associate Professor of Political Science and Public Administration, Associate Professor of Public Policy with a joint appointment in the School of Data Science at the University of North Carolina at Charlotte. His main research interests are political representation, state politics, and political methodology. 
He received his Ph.D. in Political Science from the University of North Carolina at Chapel Hill. 
 \end{IEEEbiography}

\begin{IEEEbiography}
[{\includegraphics[width=1in,height=1.25in,clip,keepaspectratio]{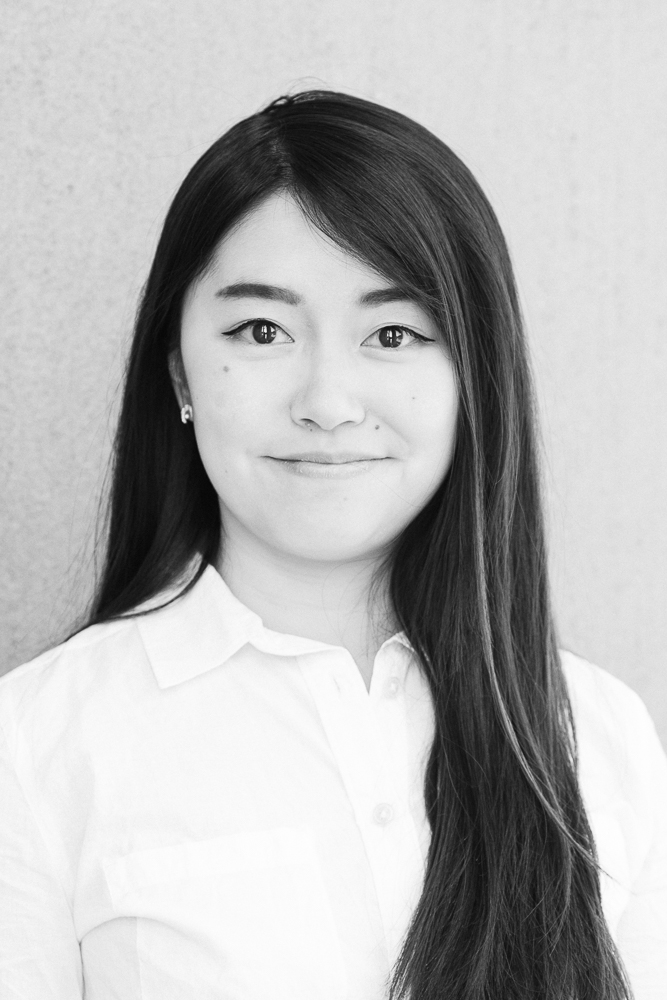}}]{Yaoyao Dai}
Yaoyao Dai is an Assistant Professor of Political Science and Public Administration at the University of North Carolina at Charlotte. She received a dual-title Ph.D. in Political Science and Asian Studies from The Pennsylvania State University. She is a computational social scientist working on information manipulation, populism, and authoritarian politics. 
 \end{IEEEbiography}

\begin{IEEEbiography}
[{\includegraphics[width=1in,height=1.25in,clip,keepaspectratio]{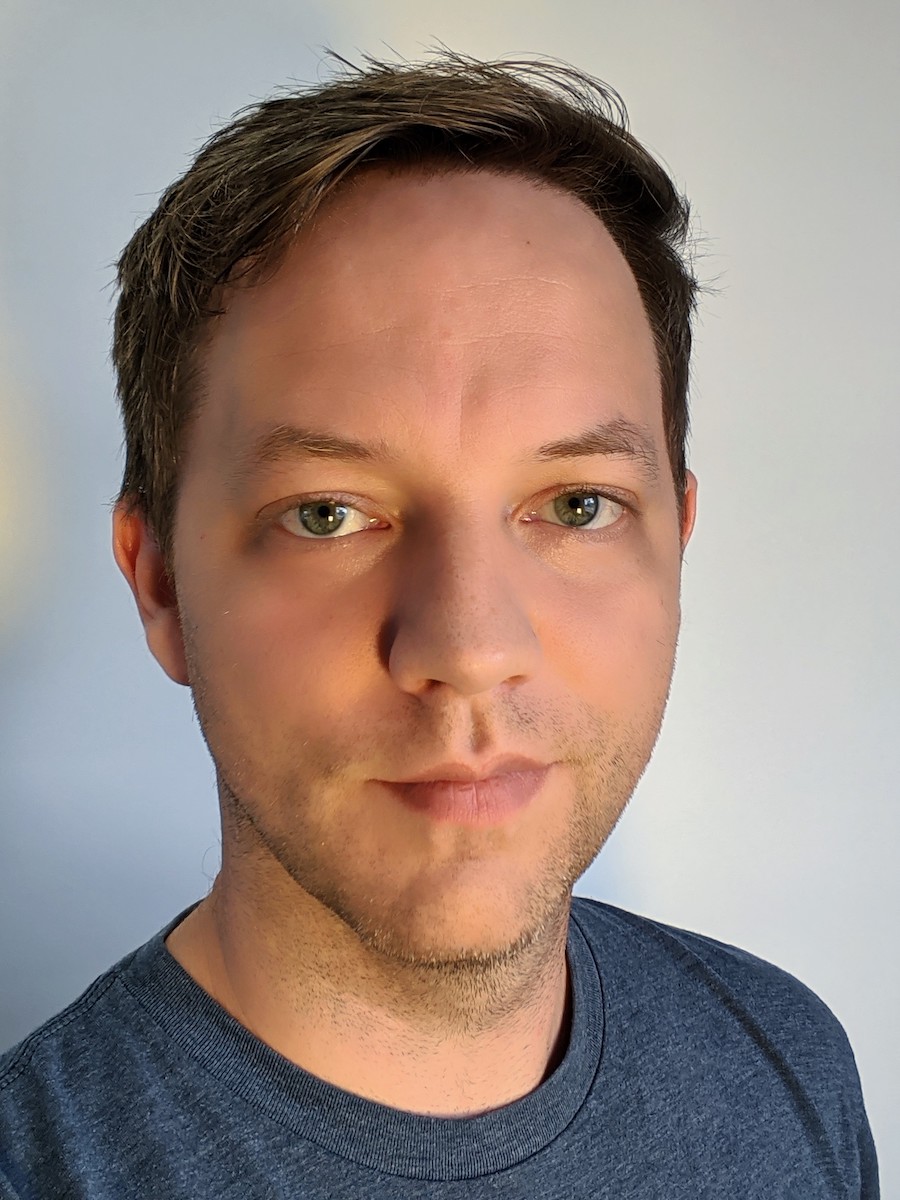}}]{Benjamin J. Radford}
Benjamin J. Radford is an Assistant Professor of Political Science and Public Administration, Assistant Professor of Public Policy, and Faculty Affiliate in the School of Data Science at the University of North Carolina at Charlotte. His research interests include political methodology, machine learning, and security, peace, \& conflict. He received his Ph.D. in Political Science from Duke University.
 \end{IEEEbiography}

\begin{IEEEbiography}[{\includegraphics[width=1in,height=1.25in,clip,keepaspectratio]{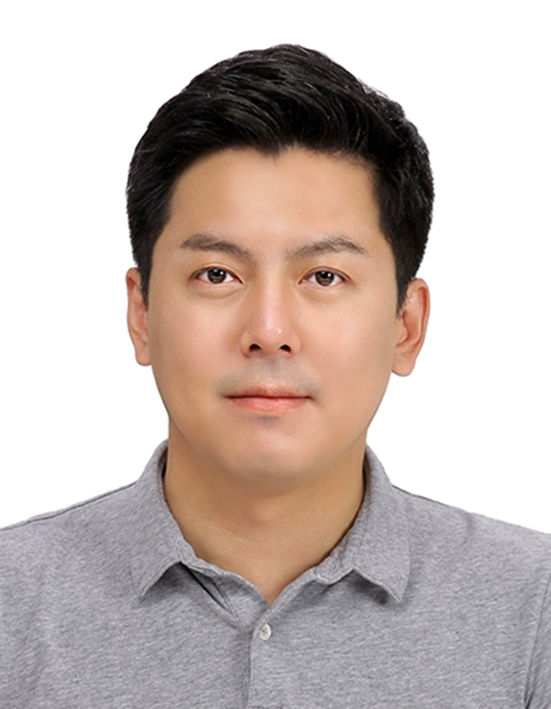}}]{Isaac Cho}
Isaac Cho is an assistant professor in the Computer Science department at Utah State University and an adjunct professor in the Computer Science department at the University of North Carolina at Charlotte. He is also a faculty member of the Ribarsky Center for Visual Analytics at the University of North Carolina at Charlotte and he directs the VizUS lab 
at Utah State University. His main research interests are interactive visual analytics, data visualization, and human-computer interactions. 
He received his Ph.D. in Computing and Information Systems from the University of North Carolina at Charlotte in 2013.
 \end{IEEEbiography}


\vfill


\end{document}